\documentclass[]{ptephy_v1}%%%%%% to generate preprint number with ptep logo

\usepackage{bm}% bold math
\begin{document}

\title{
          The general relativistic effects to the magnetic moment in the Earth's gravity.
}

\author{Takahiro Morishima}
  \affil{ 
	        Laboratory for Particle Properties, Department of Physics, Nagoya University, 
	        Furocho, Chikusa, Nagoya 464-8602, Japan %\\
  		\email{moris@nagoya-u.jp}
             }%
\author{Toshifumi Futamase}
  \affil{ 
	        Department of Astrophysics and Atmospheric Science, Kyoto Sangyo University, 
                Motoyama, Kamigamo, Kita, Kyoto 603-8555, Japan
  		\email{tof@cc.kyoto-su.ac.jp}
              }%
\author{Hirohiko M. Shimizu}
  \affil{ 
	        Laboratory for Particle Properties, Department of Physics, Nagoya University, 
	        Furocho, Chikusa, Nagoya 464-8602, Japan %\\
  		\email{hirohiko.shimizu@nagoya-u.jp}
             }%
\begin{abstract}
The magnetic moment of free fermions in the Earth's gravitational field has been studied on the basis of the general relativity.
Adopting the Schwarzschild metric for the background spacetime, the dipole coupling between the magnetic moment and the magnetic field has been found to be dependent of the gravity in the calculation up to the post-Newtonian order $O(1/c^2)$.
The gravity-dependence can be formulated by employing the effective value of the magnetic moment as a gravity-dependent quantity ${\bm \mu}_{\rm m}^{\rm eff}= (1+3\phi/c^2) \,\,{\bm \mu}_{\rm m} $ commonly for the cases of minimal coupling, non-minimal coupling and their mixture.
The gravitationally induced anomaly is found to be canceled in the experimental values of the anomalous magnetic moment measured in the Penning trap and storage ring methods.
\end{abstract}

\subjectindex{B30, B31, E00, E01}

\maketitle

\section{Introduction}
The magnetic moment is regarded as one of the fundamental properties of elementary particles.
The magnetic moment of fermions with the rest mass of $m$, the charge of $e$ and the spin of ${\bm S}\!=\!{\bm \sigma}/2$ is defined as
\begin{eqnarray}
	{\bm \mu}_{\rm m}  \equiv \frac{{\rm g}}{2}\, \frac{e}{m} {\bm S} 
	,
	\label{eq: magnetic moment.}
\end{eqnarray}
where ${\rm g}$ is the constant factor specific to each elementary particle referred to as the gyromagnetic ratio or Land$\acute{\rm e}$ g-factor
\footnote{
In this paper we use unit $\hbar$=$c$=$1$. 
}.
Here we consider the particles with the spin of 1/2 such as electrons, protons and atoms placed in the Earth's gravitational field and consider if their magnetic moment, defined by Eq.~(\ref{eq: magnetic moment.}), is independent of the gravity in the same way as the mass or the electric charge.
We focus to the influence of the spacetime distortion on the basis of the general relativity, which is not the quantum mechanical effect mediated by the graviton at the Planck scale but is the classical mechanical effect remaining even in low energy regions.

Below, we consider observables measured using instruments fixed on the Earth's surface.
The general relativity requires that the translational motion of a charged particle with the electric charge of $e$ obeys the geodesic equation
\begin{eqnarray}
           \frac{D u^\mu}{d \tau} 
      &\equiv& 
           \frac{d u^\mu}{d \tau} + \Gamma^\mu_{\nu\lambda}u^\nu u^\lambda
        =  \frac{e}{m}  F^{\mu\nu}u_\nu  
        ,
\label{eq: translational motion on curved space}
\end{eqnarray}
where $\tau$ is the proper time, $u^\mu=(u^0, {\bm u})$ is the four velocity vector, $F^{\mu\nu}$ is the electromagnetic tensor and $\Gamma^\mu_{\nu\lambda}$ is the Christoffel symbol.
The Earth's gravitational field can be described by the Schwarzschild metric, approximated up to the post-Newtonian order $O(\epsilon^2)$ in the expansion series of $\epsilon\!=\!1/c$, as
\begin{eqnarray}
ds^2 &=&    g_{\mu\nu}\, dx^{\mu}\, dx^{\nu}   \nonumber\\
     &=&    \epsilon^{-2}\Bigl( 1 \!+\! \epsilon^2 2\phi \!+\! \epsilon^4  2\phi^2  \Bigr) dt^2 
           - (1\!-\!\epsilon^2 2\phi)\,\, (dx^2 \!+\! dy^2 \!+\! dz^2)  
           \,\,\, + O(\epsilon^4)
,
\label{eq: Schwarzschild-metric}
\end{eqnarray}
where $\phi\!=\!-GM/r$ is the Earth's gravitational potential at the distance of $r\!=\!\sqrt{x^2\!+\!y^2\!+\!z^2}$ from the center of the uniformly distributed spherical mass $M$ with the radius of $R$ (See Appendix-\ref{app:Schwarzschild-geometry}).
The Eq.~(\ref{eq: translational motion on curved space}) becomes
\begin{eqnarray}
 \frac{d {\bm \beta}}{d t}  
 &=& 
    \Big( 1\!+\!(2\gamma^2\!+\!1)\epsilon^2\phi \Big) \,
    \frac{e}{\gamma\,m}\, 
       \Big(
               {\bm E} 
            + {\bm \beta}\!\times\!{\bm B} 
            - (1\!-\!4\gamma^2\epsilon^2\phi)\,{\bm \beta}\,({\bm \beta}\!\cdot\!{\bm E})
       \Big)
      \nonumber\\
 & & \hspace{10mm}
     -  \epsilon (1\!+\!\beta^2)\nabla\phi 
     +  \epsilon (4{\bm \beta}\!\cdot\!\!\nabla\phi){\bm \beta}  
      \,\,\,\,  + O(\epsilon^4)
,
\label{eq: translational motion on curved space-2}
\end{eqnarray}
where ${\bm \beta}\!=\!{\bm v}/c$, $\gamma\!=\!1/\sqrt{1\!-\!\beta^2}$ and
\begin{eqnarray}
u^0 &=& \frac{dt}{d\tau}
    = 1/\sqrt{(1+2\epsilon^2\phi+2\epsilon^4\phi^2)-(1-2\epsilon^2\phi)\beta^2} 
      \nonumber\\
    &=& \gamma \Big( 1-(2\gamma^2-1)\epsilon^2\phi \Big)  \,\,\, +O(\epsilon^4) \,\,. 
\end{eqnarray}
Here we suggest that the ${\bf E}$ and ${\bf B}$ represent the electric and magnetic fields to be measured by the observer on the Earth's surface, which correspond to the field vectors in the flat spacetime. 

At the limit of $\phi \rightarrow 0$,  the Eq.~(\ref{eq: translational motion on curved space}) results in
\begin{eqnarray}
 \frac{d {\bm \beta}}{d t}  
 &=&   
      \frac{e}{\gamma\, m}\,
          \Bigl(\! 
                   {\bm E} 
                 + {\bm \beta}\!\times\!{\bm B}  
                 - ({\bm E}\!\cdot\!{\bm \beta}) {\bm \beta}\!
          \Bigr)
,
	\label{eq: translational motion on flat space-2}
\end{eqnarray}
which is the well-known equation of motion of charged particles in the flat spacetime (Minkowski spacetime) consistently with the special relativity.
Additionally, in the region of $\beta\!=\!v/c \ll 1$, it results in the Lorentz equation of motion
\begin{eqnarray}
    \frac{d (m{\bm v})}{d t} = e ( {\bm E} + {\bm v}\!\times\!{\bm B})
.
\end{eqnarray}
For simplicity, we consider the case where the electric field is zero and the motion is limited on the horizontal plane and obtain
\begin{eqnarray}
 \frac{d {\bm \beta}}{d t}  
 &\simeq&   
   (1\!+\!3\epsilon^2\phi) \, \frac{e}{m}  {\bm \beta}\!\times\!{\bm B}  
.
\label{eq: translational motion on curved space-3}
\end{eqnarray}
According to the correspondence in the classical mechanics, the magnetic moment $\bm{\mu}_{\rm m}$ can be related to the angular momentum of a charged particle moving on a circle under a constant magnetic field, which leads to $\left\vert\bm{\mu}_{\rm m}\right\vert=e/(2m)$.
The Eq.~(\ref{eq: translational motion on curved space-3}) can be written as
\begin{eqnarray}
 \frac{d {\bm \beta}}{d t}
 \simeq   
   (1\!+\!3\epsilon^2\phi) \,\, 2\left\vert\bm{\mu}_{\rm m}\right\vert  {\bm \beta}\!\times\!{\bm B}
,
\label{eq: translational motion on curved space-4}
\end{eqnarray}
The factor $(1+3\epsilon^2\phi)$ represents the effect of the spacetime curvature due to the Earth's gravitational field.
Here we note that the $\bm{\mu}_{\rm m}$ is defined in the flat spacetime, while the magnetic field $\bm{B}$ is measured by the observer on the Earth's surface.
Therefore, in this paper, we define the effective magnetic moment as
\begin{eqnarray}
{\bm \mu}_{\rm m}^{\rm eff}  \simeq   (1\!+\!3\epsilon^2\phi) \,\, {\bm \mu}_{\rm m}   ,
\end{eqnarray}
to include the gravitational effect.
Above discussion remains within the classical picture since the lepton is not treated as the fermion with the quantized spin and it is not trivial to extend the influence of the gravity for quantum mechanical particles.
Therefore the effective magnetic moment should be examined based on the formalism compatible to both the quantum mechanics and the general relativity.

In the quantum field theory, the magnetic moment of a fermion is related to the form factors $F_1$ and $F_2$ defined in the vertex function
(See \cite{itzykson-1985,peskin-1997})
\begin{eqnarray}
 {\cal M}^\mu (p',p) =   
	\gamma^\mu F_1(q^2) 
	+
	\frac{i}{2m} \sigma^{\mu\nu}q_\nu F_2(q^2)
	,
\label{eq: vertex function.}
\end{eqnarray}
and the g-factor in the flat spacetime is defined as
\begin{eqnarray}
  {\rm g} 
  = 2\,(F_1(0)+F_2(0)) 
  .
\label{eq: magnetic moment. 2}
\end{eqnarray}
The origin of the magnetic moment varies according to the type of fermions as  
$F_1(0)\!=\!1$, $F_2(0)\!=\!0$ for Dirac particles, {\it i.e.,} fermions with g=2 for which their internal structure and radiative corrections can be ignored,
$F_1(0)\!=\!0$, $F_2(0)\!\ne\!0$ for neutrons and neutral atoms, {\it i.e.,} fermions with g$\ne$2 and e=0,
and
$F_1(0)\!=\!1$, $F_2(0)\!\ne\!0$ for protons and ions, {\it i.e.,} fermions with g$\ne$2,\,e$\ne$0.
We applied the quantum mechanics in the curved spacetime in the general relativity for three types:
\begin{itemize}
 \item[] \hspace{7mm} {\bf Type-A:}\,\,\,\,\, Dirac particles with g=2 and e$\ne$0  \hspace{3mm}  ($F_1(0)=1$,\, $F_2(0)=0 $),
 \item[] \hspace{7mm} {\bf Type-B:}\,\,\,\,\, fermions with g$\ne$2 and e=0         \hspace{13mm} ($F_1(0)=0$,\, $F_2(0)\ne$ 0),
 \item[] \hspace{7mm} {\bf Type-C:}\,\,\,\,\, fermions with g$\ne$2 and e$\ne$0     \hspace{13mm} ($F_1(0)=1$,\, $F_2(0)\ne$ 0).
\end{itemize}
The coupling between the fermions and the electromagnetic field corresponds to 
the minimal coupling for Type-A,
the non-minimal coupling for Type-B
and 
the mixture of minimal and non-minimal couplings for Type-C.

In this paper,
we generalize the Dirac equation in curved spacetime by employing the covariant formulation.
For simplicity, we employ the Schwarzschild metric as the background spacetime, derive the hamiltonian of free fermions to be observed using instruments fixed on the Earth's surface 
up to the post-Newtonian order $O(1/c^2)$ for above three types and evaluated the effective value of the magnetic moment.
We introduce the {\it effective} magnetic moment of fermions in the Earth's gravitational field and discuss the influence to precision measurements of the anomalous magnetic moment of electrons and muons.

\section{Effective Magnetic Moment in Curved Spacetime}
The influence of the general relativity in the hamiltonian of fermions has been considered and reported in several papers~\cite{anandan-1984,parker-1980,wajima-1997}.
Wajima {\it et al.}~\cite{wajima-1997} employed the covariant Dirac equation
\begin{equation}
  \Bigl(  \gamma^\mu (i\nabla_\mu ) -  \epsilon^{-1} m \Bigr) \Psi=0
  \label{eq:Dirac-covariant-without-e}
\end{equation}
and regarded its time component
%\begin{widetext}
\begin{eqnarray}
 {\cal H}_{} \!\!
      &\equiv& \!\!  i\partial_0  \nonumber\\
      &=& \!\!\! - i \Gamma_0 \!+\! \frac{1}{g^{00}}\gamma^0\gamma^i (-i\partial_i \!-\! i \Gamma_i) 
           \!+\! \epsilon^{-1}\frac{1}{g^{00}}\gamma^0 m 
  \,\,\,\,\,\,\,\,
  \label{eq:Hamiltonian-covariant-without-e}
\end{eqnarray}
as the hamiltonian and calculated its expectation value up to the post-Newtonian order $O(1/c^2)$,
where $\Psi$ is the spinor function,
$\nabla_\mu = \partial_\mu + \Gamma_\mu$ the covariant derivative,
$\Gamma_\mu$ the spin connection,
$\gamma^\mu$ the $\gamma$-matrices in the curved spacetime (Appendix-\ref{app:gamma}),
and $\epsilon\!=\!1/c$ the small quantity to be used as the variable for the power series expansion to identify the post-Newtonian approximation.
Since the electric charge and the magnetic moment were not involved in their method, we develop and generalize their method to derive the hamiltonian containing gravitational effects in the general relativity on the basis of the covariant Dirac equation with the electric charge and the magnetic moment.
Here we adopt the Heisenberg equation for the magnetic moment $\bm{\mu}_{\rm m}$
\begin{eqnarray}
    \frac{d{\bm S}}{dt} 
 =  \frac{1}{i} \Bigl[ {\bm S},  {\cal H}  \Bigr] 
 =  {\bm \mu}_{\rm m} \times {\bm B}
\label{eq: Heisenberg eq.}
\end{eqnarray}
as the definition of the magnetic moment in the flat spacetime.
We generalize the Eq.~(\ref{eq: Heisenberg eq.}) using the general relativistic hamiltonian to evaluate the effective magnetic moment $\bm{\mu}_{\rm m}^{\rm eff}$.

\subsection{Dirac Particles with g=2 (Type-A)}  % Type-A
In this section, we consider the Type-A, for which the form factors are $F_1(0)\!=\!1$ and $F_2(0)\!=\!0$, corresponding to point-like fermions,  {\it i.e.},  electrons and muons without radiative corrections.
The covariant Dirac equation for this case is given with the minimal coupling with the four potential $A_{\mu}$ as
\begin{eqnarray}
  \Bigl( \,\gamma^\mu\, i\,(\nabla_\mu + i\, e A_\mu )  -  \epsilon^{-1} m \, \Bigr) \,\,\Psi=0 
,
  \label{eq: Dirac eq. of lepton on curved space}
\end{eqnarray}
which leads to the hamiltonian
\begin{eqnarray}
{\cal H}^{^{\rm A}}    
 &=&  
    - i\, \Gamma_0 \!+\! e A_0
     \!+\! \frac{1}{g^{00}}\gamma^0\gamma^i (-i\,\partial_i \!-\! i\,\Gamma_i \!+\! e  A_i) 
     + \epsilon^{-1} \frac{1}{g^{00}}\gamma^0 m
,
\label{eq: Hamiltonian of lepton on curved space}
\end{eqnarray}
which is the general form of the hamiltonian of Dirac particles with g=2 with the electric charge and the magnetic moment.
It is uniquely determined for a given metric of the spacetime.
Adopting the Schwarzschild metric Eq.~(\ref{eq: Schwarzschild-metric}), we can rewrite Eq.~(\ref{eq: Hamiltonian of lepton on curved space}), after some algebra to rearrange (See Appendix-\ref{app:re-normalization}), as
\begin{eqnarray}
{\cal H}^{^{\rm A}}
    &=&   \epsilon^{-2} \rho_3 m 
          + \epsilon^{-1} \rho_1  {\bm \sigma}\!\cdot\!( {\bm p} \!-\! e {\bm A}) 
        + e A_0
        + \rho_3  m\phi 
        + \epsilon\, \rho_1
              \Bigl\{ 
                      {\bm \sigma}\!\cdot\!( {\bm p} \!-\! e {\bm A}),\, \phi
              \Bigr\}
        + \epsilon^2 
                \rho_3 \frac{m\phi^2}{2} 
          \nonumber\\
     & &
          \,\,\,\,\,+ O(\epsilon^4)
.
\label{eq: Hamiltonian of lepton written by Schwarzschild metric}
\end{eqnarray}
This hamiltonian is a $4\times4$ matrix.
We can obtain a $2\times2$ even matrix by applying the Foldy-Wouthuysen-Tani transformation~\cite{FWT}.
Its left-up $2\times2$ matrix (the large component) ${\cal H}_{+}^{^{\rm A}}$ given as
\begin{eqnarray}
{\cal H}_{+}^{^{\rm A}}   
    &=&   \epsilon^{-2} m 
          \nonumber\\
    & & +  \frac{({\bm \sigma}\!\cdot\!({\bm p}\!-\!e{\bm A}))^2}{2m} + m\phi 
            + e A_0
          \nonumber\\
    & & 
        + \epsilon^2 
          \Bigg( 
                   \frac{m\phi^2}{2} 
                  - \frac{({\bm \sigma}\!\cdot\! ({\bm p}\!-\!e{\bm A}))^4  }{8m^3}
                  + \frac{3}{8m} 
                      \Big\{ {\bm \sigma}\!\cdot\!({\bm p}\!-\!e{\bm A}),\, 
                      \big\{{\bm \sigma}\!\cdot\!({\bm p}\!-\!e{\bm A}),\,\phi \big\}\Big\}  
          \nonumber\\
    & & \hspace{0.7cm}
         - \frac{1}{8m^2}
           \Big[ {\bm \sigma}\!\!\cdot\!({\bm p}\!-\!e{\bm A}),  \big[{\bm \sigma}\!\!\cdot\!({\bm p}\!-\!e{\bm A}), \!e A_0 \big] \Big] 
          \Bigg)
            \hspace{0.5cm}
           + O(\epsilon^4)
\label{eq: Effective Hamiltonian of lepton 1}
\end{eqnarray}
corresponds to the hamiltonian of the Dirac particle in curved spacetime.
Substituting ${\bm \mu}_{\rm m}=e{\bm S}/m=e{\bm \sigma}/2m$, Eq.~(\ref{eq: Effective Hamiltonian of lepton 1}) leads to
\begin{eqnarray}
{\cal H}_{+}^{^{\rm A}}   
      &=&    \epsilon^{-2}\,m  \nonumber\\
      & &  + 
                    \frac{({\bm p} \!-\! e {\bm A})^2}{2m}  
                   - {\bm \mu}_{\rm m}\!\cdot\!{\bm B}
                   + m\phi
                   + e A_0 
            \nonumber\\
      & & 
        + \epsilon^2 
          \Bigg( 
                   \frac{m\phi^2}{2} 
                   - \frac{({\bm p} - e {\bm A})^4}{8m^3}
                   - \frac{({\bm \mu}_{\rm m}\!\cdot\!{\bm B})^2}{2m}
                   - \frac{1}{4m^2} \nabla^2({\bm \mu}_{\rm m}\!\cdot\!{\bm B}) 
                   - 3\phi\,\, {\bm \mu}_{\rm m}\!\cdot\!{\bm B}
                  \nonumber\\
      & & \hspace{0.7cm}
                   + \frac{1}{2m^2}
                        ({\bm p}\!-\!e {\bm A}) \!\cdot\!\Bigl({\bm \mu}_{\rm m}\!\cdot\!{\bm B}\, ({\bm p}\!-\!e {\bm A})\Bigr)
                  + \frac{3}{2m} 
                         ({\bm p}\!-\!e{\bm A})\!\cdot\!\Bigl(\phi({\bm p}\!-\!e{\bm A}) \Bigr)
                  \nonumber\\
      & & \hspace{0.7cm}
                  - \frac{3\phi}{2m r^2}{\bm S}\!\cdot\!\Big({\bm r}\!\times\!({\bm p}\!-\!e{\bm A})\Big)
          \nonumber\\
      & & \hspace{0.7cm}
         - \frac{{\mu}_{\rm m}}{4m}\,\,  \nabla\cdot({\bm E}+\frac{\partial{\bm A}}{\partial t} )
         - \frac{1}{2m}{\bm \mu}_{\rm m}\!\cdot\!\Big(({\bm E}+\frac{\partial {\bm A}}{\partial t})\!\times\!({\bm p}\!-\!e{\bm A})\Big)
        \Bigg) 
      \,\,\,   + O(\epsilon^4)
.
\label{eq: Effective Hamiltonian of lepton 2}
\end{eqnarray}
Terms at the order of $\epsilon^{-2}$ and $\epsilon^{0}$ are well known, while some of terms at the order of $\epsilon^{2}$ are newly introduced coupling terms between the gravitational potential $\phi$, the magnetic moment ${\bm \mu}_{\rm m}$ and the magnetic field ${\bm B}$.
The effective magnetic moment is given as
\begin{eqnarray}
&&  {\bm \mu}_{\rm m}^{\rm eff} \simeq \,  (1+ 3\epsilon^2\phi) \,\,{\bm \mu}_{\rm m}
    \hspace{1cm}
   (\mbox{for Type-A})
\label{eq: Effective magnetic moment of lepton}
\end{eqnarray}
for sufficiently slow particles moving in a static spacetime.
The result is consistent with {\it the gravitationally modified magnetic moment} obtained in the $TH\epsilon\mu$ formalism~\cite{lightman-1973,will-1974}
\footnote{
The $TH\epsilon\mu$ formalism with the Schwarzschild metric corresponds to ${T\!=\!1\!+\!\epsilon^2 2\phi\!+\!\epsilon^4 2\phi^2}$ and ${H\!=\!1\!-\!\epsilon^2 2\phi}$, which leads to the gravitational dependence of ${T^{1/2}/H\!=\!1\!+\!3\epsilon^2\phi}$\,\, for the magnetic moment~\cite{will-1974}.
}.
The above discussions are restricted to Dirac fermions minimally coupled to electromagnetic fields.
Non-minimal coupling case such as neutrons and spatially spread case such as fermions dressed with the renormalization and quantum radiative corrections are not involved.

\subsection{Fermions with g$\ne$2, e=0 (Type-B)}   % Type-B
Neutrons and neutral atoms corresponds to the case of form factors of $F_1(0)\!=\!0$ and $F_2(0)\!\ne\!0$ since e=0.
Since they cannot minimally coupled to the electromagnetic field, their covariant Dirac equation contains the following non-minimal coupling between the magnetic moment $\mu_{\rm m}$ and the electromagnetic tensor $F_{\mu\nu}$
\begin{eqnarray}
   \Big(   \gamma^\mu \,i\, \nabla_\mu   - \epsilon^{-1} m
 - \epsilon\,\frac{1}{2}{\mu}_{\rm m}\,\sigma^{\mu\nu} F_{\mu\nu} \Big) \Psi=0 
,
\label{eq: Dirac eq. of neutron on curved space}
\end{eqnarray}
where
$\sigma^{\mu\nu}$ and $F_{\mu\nu}$ are 2nd-rank covariant tensor (see  Appendix-\ref{app:gamma}).
The hamiltonian is calculated as
\begin{eqnarray}
{\cal H}^{^{\rm B}}    
 &=&  - i\, \Gamma_0 + \frac{1}{g^{00}}\gamma^0\gamma^i (-i\,\partial_i -  i\,\Gamma_i) 
      + \epsilon^{-1}\frac{1}{g^{00}}\gamma^0 m (1+\epsilon^2\,\frac{\mu_{\rm m}}{2m}\,\sigma^{\mu\nu} F_{\mu\nu} )
.
\label{eq: Hamiltonian of neutron on curved space}
\end{eqnarray}
In the same manner described in the previous section, the large component in the Schwarzschild spacetime ${\cal H}_{+}^{^{\rm B}}$ is obtained as
\begin{eqnarray}
{\cal H}_{+}^{^{\rm B}}   
      &=&    \epsilon^{-2} m \nonumber\\
      & &   
           +  \frac{({\bm \sigma}\!\cdot\!{\bm p})^2}{2m} 
                  + m\phi
                  - {\bm \mu}_{\rm m}\!\cdot\!{\bm B} 
             \nonumber\\
      & &  + \epsilon^2 \Bigg( 
                      \frac{m\phi^2}{2} 
                    - \frac{({\bm \sigma}\!\cdot\!{\bm p})^4}{8m^3}
                    - 3\,\phi\,{\bm \mu}_{\rm m}\!\cdot\!{\bm B} 
           + \frac{3}{8m}\big\{{\bm \sigma}\!\cdot\!{\bm p}, \, \{{\bm \sigma}\!\cdot\!{\bm p},\,\phi\}  \big\} 
           \nonumber\\
      & &    \hspace{10mm} 
           + \frac{1}{8m^2}\Big\{{\bm \sigma}\!\cdot\!{\bm p}, \,\{{\bm \sigma}\!\cdot\!{\bm p},\,{\bm \mu}_{\rm m}\!\cdot\!{\bm B} \} \Big\}    
                        \Bigg) 
           \nonumber\\
      & &   
         - \epsilon^2 \frac{i}{2m} \big[ {\bm \sigma}\!\cdot\!{\bm p},\,\, {\bm \mu}_{\rm m}\!\cdot\!{\bm E}\big] 
         + O(\epsilon^4)
\nonumber\\
\nonumber\\
    &=&  
          \epsilon^{-2}  m           
          \nonumber\\
    & & 
        + \frac{{\bm p}^2 }{2 m}  + m\phi - {\bm \mu}_{\rm m}\!\cdot\!{\bm B} 
          \nonumber\\
    & & 
        + \epsilon^2 
          \Bigg(  
                  \frac{m\phi^2}{2}
                - \frac{{\bm p}^4  }{8 m^3}  
                - 3\phi \,{\bm \mu}_{\rm m}\!\cdot\!{\bm B} 
                + \frac{3}{2m}\, {\bm p}\!\cdot\!(\phi\,{\bm p}) -  \frac{3\phi}{2m r^2} {\bm S}\!\cdot\!({\bm r}\!\times\!{\bm p})    
          \nonumber\\
    & & 
          \hspace{10mm} 
                 - \frac{1}{8m^2} \nabla^2 ({\bm \mu}_{\rm m}\!\cdot\!{\bm B}) 
                 + \frac{1}{4m^2}({\bm \mu}_{\rm m}\!\cdot\!{\bm p})\, ({\bm B}\!\cdot\!{\bm p} )
          \nonumber\\
    & & 
          \hspace{10mm} 
                 + \frac{1}{4m^2} ({\bm B}\!\cdot\!{\bm p}) \, ({\bm \mu}_{\rm m}\!\cdot\!{\bm p})
                 + \frac{\mu_{\rm m}}{4m^2}  (\nabla\!\!\times\!\!{\bm B})\!\times\!{\bm p} 
          \nonumber\\
    & & 
          \hspace{10mm} 
                 - \frac{1}{2m} {\mu}_{\rm m}\, \nabla\!\cdot\!{\bm E} 
                 - \frac{1}{m}  {\bm \mu}_{\rm m}\!\cdot\!({\bm E}\!\times\!{\bm p})   
                 - \frac{i}{2m}\, {\bm \mu}_{\rm m}\!\cdot\! (\nabla\!\times\!{\bm E})  
          \Bigg)
        \,\,\, + O(\epsilon^4) 
        .
\label{eq: Effective Hamiltonian of neutron}
\end{eqnarray}
Accordingly, the effective magnetic moment is obtained as
\begin{eqnarray}
 {\bm \mu}_{\rm m}^{\rm eff} \simeq (1+ 3\,\epsilon^2\,\phi) \,\,{\bm \mu}_{\rm m} 
 \hspace{1cm}
 (\mbox{for Type-B})
\label{eq: Effective magnetic moment of neutron}
\end{eqnarray}
This result shows that the effective magnetic moment with the general relativistic influence up to the order of $O(\epsilon^2)$ has the common form both for Type-A and Type-B, in which the minimal and non-minimal coupling terms are employed in the Dirac equations Eq.~(\ref{eq: Dirac eq. of lepton on curved space}) and Eq.~(\ref{eq: Dirac eq. of neutron on curved space}), respectively.

\subsection{Fermions with g$\ne$2, e$\ne$0 (Type-C)} % type-C
Spread charged fermions have the form factors of $F_1(0)\!=\!1$ and $F_2(0)\!\ne\!0$.
This case corresponds to electrons and muons dressed with quantum radiative corrections and also composite particles such as protons and atoms containing internal structure.
The corresponding covariant Dirac equation contains both the minimal coupling and non-minimal coupling terms as
\begin{eqnarray}
   \Big(   
         \gamma^\mu \,i\,(\nabla_\mu  + i\,e A_\mu)  - \epsilon^{-1} m
         - \epsilon\,\,\frac{{\rm g}\!-\!2}{2}\frac{e}{4m}\sigma^{\mu\nu}F_{\mu\nu}  
   \Big) \Psi=0 
.
\label{eq: Dirac eq. of proton on curved space}
\end{eqnarray}
The influence of radiative corrections and internal structure are contained in the last term in Eq.~(\ref{eq: Dirac eq. of proton on curved space}), which can be confirmed to observe that it returns to Eq.~(\ref{eq: Dirac eq. of lepton on curved space}) to eliminate the last term corresponding to the spread of fermions.
The hamiltonian in $4\times4$-matrix format is calculated as
\begin{eqnarray}
{\cal H}^{\rm C} 
 &=&  - i\, \Gamma_0 + e A_0
    + \frac{1}{g^{00}}\gamma^0\gamma^i (-i\,\partial_i - i\,\Gamma_i + e  A_i) 
    + \epsilon^{-1} \frac{1}{g^{00}}\gamma^0 m
      {\Big(1 + \epsilon^2\,\frac{{\rm g}\!-\!2}{2}\frac{e}{4m^2}\sigma^{\mu\nu}F_{\mu\nu}\Big)}  
,
      \nonumber\\
\label{eq: Hamiltonian of proton on curved space}
\end{eqnarray}
and the large component in $2\times2$-matrix format hamiltonian ${\cal H}_{+}^{^{\rm C}}$ as
\begin{eqnarray}
{\cal H}_{+}^{^{\rm C}}   
    &=&   \epsilon^{-2}  m 
          \nonumber\\
    & & +  \frac{({\bm \sigma}\!\cdot\!({\bm p}\!-\!e{\bm A}))^2}{2m} + m\phi 
            + e A_0
            - {\it \Delta}{\bm \mu}_0\!\cdot\!{\bm B} 
          \nonumber\\
    & & 
        + \epsilon^2 
          \Bigg( 
                   \frac{m\phi^2}{2} 
                  - \frac{({\bm \sigma}\!\cdot\! ({\bm p}\!-\!e{\bm A}))^4  }{8m^3}
                  - 3\,\phi\,{\it \Delta}{\bm \mu}_0\!\cdot\!{\bm B} 
          \nonumber\\
    & & \hspace{0.7cm}
                  + \frac{3}{8m} 
                      \Big\{ {\bm \sigma}\!\cdot\!({\bm p}\!-\!e{\bm A}),\, 
                      \big\{{\bm \sigma}\!\cdot\!({\bm p}\!-\!e{\bm A}),\,\phi \big\}\Big\}  
         - \frac{1}{8m^2}
           \Big[ {\bm \sigma}\!\!\cdot\!({\bm p}\!-\!e{\bm A}),  \big[{\bm \sigma}\!\!\cdot\!({\bm p}\!-\!e{\bm A}), \!e A_0 \big] \Big] 
           \nonumber\\
    & &  \hspace{0.7cm}
           + \frac{1}{8m^2}\Big\{{\bm \sigma}\!\cdot\!({\bm p}\!-\!e{\bm A}), \,\{{\bm \sigma}\!\cdot\!({\bm p}\!-\!e{\bm A}),\,{\it \Delta}{\bm \mu}_0\!\cdot\!{\bm B} \} \Big\}    
                        \Bigg) 
           \nonumber\\
    & &   
            - \epsilon^2 \frac{i}{2m} \big[ {\bm \sigma}\!\cdot\!({\bm p}\!-\!e{\bm A}), {\it \Delta}{\bm \mu}_0\!\cdot\!{\bm E}\big] 
         + O(\epsilon^4)
          \nonumber\\
          \nonumber\\
    &=&   
          \epsilon^{-2} m 
          \nonumber\\
    & & 
        + \frac{ ({\bm p} \!-\! e {\bm A})^2 }{2m} 
        + m\phi  
        + e A_0  
        - ( {\bm \mu}_0 \!+\!  \Delta{\bm \mu}_0 ) \!\cdot\! {\bm B} 
          \nonumber\\
    & & 
        + \epsilon^2 
          \Bigg( 
                   \frac{m\phi^2}{2}
                 - \frac{({\bm p} \!-\! e {\bm A})^4}{8 m^3} 
                 - 3 \phi\,\, ( {\bm \mu}_0 \!+\!  \Delta{\bm \mu}_0 ) \!\cdot\! {\bm B} 
                   \nonumber\\
    & &          \hspace{7mm}
                 - \frac{({\bm \mu}_0\!\cdot\!{\bm B})^2}{2 m} 
                 - \frac{1}{4 m^2} \nabla^2 ({\bm \mu}_0\!\cdot\!{\bm B}) 
                 + \frac{1}{2 m^2} ({\bm p}\!-\!e {\bm A})\!\cdot\! \Big( {\bm \mu}_0\!\cdot\!{\bm B}\,\,({\bm p}\!-\!e {\bm A}) \Big)
                   \nonumber\\
    & &          \hspace{7mm}
                 + {\frac{3}{2m}}  ({\bm p}\!-\!e{\bm A}) \!\cdot\!\Big( \phi({\bm p}\!-\!e{\bm A}) \Big)
                 - {\frac{3\phi}{2m r^2}}  {\bm S}\!\cdot\!\Big({\bm r}\!\times\!({\bm p}\!-\!e{\bm A})\Big) 
                   \nonumber\\
    & & 
                 \hspace{7mm}
                 - \frac{1}{8m^2} \nabla^2 (\Delta{\bm \mu}_0\!\cdot\!{\bm B}) 
                 + \frac{1}{4m^2} \Big(\Delta{\bm \mu}_0\!\cdot\!({\bm p}\!-\!e{\bm A})\Big)\,\, \Big({\bm B}\!\cdot\!({\bm p}\!-\!e{\bm A})\Big)
                   \nonumber\\
    & & 
                 \hspace{7mm}
                 + \frac{1}{4m^2} \Big({\bm B}\!\cdot\!({\bm p}\!-\!e{\bm A})\Big) \,\, \Big(\Delta{\bm \mu}_0\!\cdot\!({\bm p}\!-\!e{\bm A})\Big)
                 + \frac{\Delta\mu_0}{4m^2}\, (\nabla\!\times\!{\bm B})\!\times\!{\bm p}
                   \nonumber\\
    & &   
                 \hspace{7mm}
                  - \frac{\mu_0}{4m} \,\nabla\cdot ({\bm E}\!+\!\frac{\partial {\bm A}}{\partial t} ) 
                  - \frac{1}{2m} {\bm \mu}_0\!\cdot\!\Big(({\bm E}\!+\!\frac{\partial {\bm A}}{\partial t})\!\times\! ({\bm p} \!-\! e {\bm A})\Big)
                    \nonumber\\
    & & 
                 \hspace{7mm}
                  - \frac{\Delta \mu_0}{2m}  \nabla\!\cdot\!{\bm E} 
                  - \frac{1}{m}  \Delta{\bm \mu}_0\!\cdot\!({\bm E}\!\times\!{\bm p})  
                  - \frac{i}{2m}  \Delta{\bm \mu}_0\!\cdot\!(\nabla\!\times\!{\bm E}) 
          \Bigg)
                  \,\,\, + O(\epsilon^4) 
	,
\end{eqnarray}
where
\begin{eqnarray}
{\bm \mu}_0             &\equiv&  \frac{e}{m}{\bm S} = \frac{e}{2m}{\bm \sigma}  ,    \hspace{5mm}\mbox{(Bohr magneton)}\nonumber\\
\Delta{\bm \mu}_0       &\equiv&  {\bm \mu}_m - {\bm \mu}_0 = \frac{{\rm g}-2}{2}\frac{e}{2m}{\bm \sigma}
\end{eqnarray}
Accordingly, the effective magnetic moment is
\begin{eqnarray}
 {\bm \mu}_{\rm m}^{\rm eff} 
&\simeq&
    ( 1 + 3\epsilon^2\phi) \,\, {\bm \mu}_{\rm m} 
 \hspace{1cm}
 (\mbox{for Type-C})
\label{eq: Effective magnetic moment of proton}
\end{eqnarray}
The Eq.~(\ref{eq: Effective magnetic moment of proton}) can be understood as the general expression of the effective magnetic moment of fermions including leptons and hadrons.
Although the g-factor of proton or atoms largely deviates from g=2, the effective magnetic moment is given in the same form as Type-A : Dirac particles with minimal coupling Eq.~(\ref{eq: Dirac eq. of lepton on curved space}) and Type-B: neutral fermions with non-minimal coupling Eq.~(\ref{eq: Dirac eq. of neutron on curved space}).

\subsection{Effective Anomalous Magnetic Moment in Curved Spacetime}
The effective magnetic moment to be observed with the instrument fixed on the Earth's surface has been obtained as Eq.~(\ref{eq: Effective magnetic moment of proton}) up to the post-Newtonian order $O(1/c^2)$ adopting the Schwarzschild metric as the background spacetime, common to leptons and hadrons of Type-A, Type-B and Type-C.
The effective values of the gyromagnetic ratio (g-factor) and the anomalous magnetic moment ${a\equiv{\rm g}/2\!-\!1={\mu}_{\rm m}/{\mu}_0-1}$, to be measured in ground-based experiments, can be defined as
\begin{eqnarray}
{\rm g}^{\rm eff}
    &\simeq& (1 + 3\,\epsilon^2\,\phi) \,\, {\rm g} 
,
             \nonumber\\
{a}^{\rm eff} 
    &\simeq& a \, + 3\, (1+a)\epsilon^2\phi 
,
\end{eqnarray}
which implies that the ${\rm g}$-factor and the anomalous magnetic moment depends on the gravitational field as the result of the influence of the curved spacetime in the general relativity.
For an example, the anomalous magnetic moment of fermions with g$\simeq$2 is evaluated to be measured as 
$|a^{\rm eff}| \simeq |3\epsilon^2\phi| \simeq 2.1\times 10^{-9}$ 
on the Earth's surface.

It implies that the Earth's gravitational field induces the anomalous magnetic moment of $2.1\!\times\! 10^{-9}$ for electrons, muons, {\it etc.} regardless of the type of fermions, in addition to quantum radiative corrections.

\subsection{Parametric Post-Newtonian (PPN) Approximation}
Here we consider the case in which the Parametric Post-Newtonian (PPN) approximation is adopted as the background spacetime:
\begin{eqnarray}
\!\!\!\!\!\!\!
ds^2 &=&   \epsilon^{-2} \Big( 1 +\epsilon^2 2\phi +\epsilon^4 2 \beta_* \phi^2 \Big) dt^2  
           - (1-\epsilon^2 2\gamma_* \phi)\,\, (dx^2 + dy^2 + dz^2)  
.
\label{eq: PPN-metric}
\end{eqnarray}
This is the primitive version of one of the possible extension of the general relativity~\cite{PPN}, in which the parameter choice of $\beta_* \ne 1$ and $\gamma_* \ne 1$ corresponds to the Scalar-Tensor gravity model and the choice of $\beta_* = 1$, $\gamma_* = 1$ corresponds to the standard general relativity.
In the same manner as the case of general relativity, the large component hamiltonian can be obtained via the covariant Dirac equation Eq.~(\ref{eq: Dirac eq. of proton on curved space}) using the PPN metric as
\begin{eqnarray}
{\cal H}^{\rm PPN}_{+} 
    &=&   
          \epsilon^{-2} m 
          \nonumber\\
    & & + e A_0  
        + m\phi  
        + \frac{ ({\bm p} \!-\! e {\bm A})^2 }{2m} 
        - ( {\bm \mu}_0 \!+\!  \Delta{\bm \mu}_0 ) \!\cdot\! {\bm B} 
          \nonumber\\
    & & 
        + \epsilon^2 
          \Bigg( 
                   (\beta^*\!\!-\!\frac{1}{2}) m\phi^2
                 - (1\!+\!2\gamma^*) \phi\,\, ( {\bm \mu}_0 \!+\!  \Delta{\bm \mu}_0 ) \!\cdot\! {\bm B} 
                   \nonumber\\
    & &          \hspace{7mm}
                 - \frac{({\bm p} \!-\! e {\bm A})^4}{8 m^3} 
                 - \frac{({\bm \mu}_0\!\cdot\!{\bm B})^2}{2 m} 
                 - \frac{1}{4 m^2} \nabla^2 ({\bm \mu}_0\!\cdot\!{\bm B}) 
                   \nonumber\\
    & &          \hspace{7mm}
                 + \frac{1}{2 m^2} ({\bm p}\!-\!e {\bm A})\!\cdot\! \Big( {\bm \mu}_0\!\cdot\!{\bm B}\,\,({\bm p}\!-\!e {\bm A}) \Big)
                   \nonumber\\
    & &         \hspace{7mm}
                 + \frac{(1\!+\!2\gamma^*)}{2m} ({\bm p}-e{\bm A}) \!\cdot\!\Big( \phi({\bm p}-e{\bm A}) \Big)
                 - \frac{(1\!+\!2\gamma^*)\phi}{2m r^2} {\bm S}\!\cdot\!\Big( {\bm r}\!\times\!({\bm p}-e{\bm A})\Big) 
                   \nonumber\\
    & & 
                 \hspace{7mm}
                 - \frac{1}{8m^2} \nabla^2 (\Delta{\bm \mu}_0\!\cdot\!{\bm B}) 
                 + \frac{\Delta\mu_0}{4m^2}\, (\nabla\!\times\!{\bm B})\!\times\!{\bm p}
                   \nonumber\\
    & & 
                 \hspace{7mm}
                 + \frac{1}{4m^2} \Big(\Delta{\bm \mu}_0\!\cdot\!({\bm p}\!-\!e{\bm A})\Big)\, \Big({\bm B}\!\cdot\!({\bm p}\!-\!e{\bm A})\Big)
                 + \frac{1}{4m^2} \Big({\bm B}\!\cdot\!({\bm p}\!-\!e{\bm A})\Big) \, \Big(\Delta{\bm \mu}_0\!\cdot\!({\bm p}\!-\!e{\bm A})\Big)
                   \nonumber\\
    & &   
                 \hspace{7mm}
                  - \frac{\mu_0}{4m} \,\nabla\cdot ({\bm E}\!+\!\frac{\partial {\bm A}}{\partial t} ) 
                  - \frac{1}{2m} {\bm \mu}_0\!\cdot\!\Big(({\bm E}\!+\!\frac{\partial {\bm A}}{\partial t})\!\times\! ({\bm p} \!-\! e {\bm A})\Big)
                    \nonumber\\
    & & 
                 \hspace{7mm}
                  - \frac{\Delta \mu_0}{2m}  \nabla\!\cdot\!{\bm E} 
                  - \frac{1}{m}  \Delta{\bm \mu}_0\!\cdot\!({\bm E}\!\times\!{\bm p})  
                  - \frac{i}{2m}  \Delta{\bm \mu}_0\!\cdot\!(\nabla\!\times\!{\bm E}) 
          \Bigg)
                  + O(\epsilon^4) 
.
\end{eqnarray}
Accordingly, the effective magnetic moment can be derived, in the same manner, as
\begin{eqnarray}
 {\bm \mu}_{\rm m}^{\rm eff} 
&\simeq&
    \Big( 1 + (1+2\gamma^*)\,\epsilon^2\phi \Big) \,\, {\bm \mu}_{\rm m} 
 \hspace{7mm}
 (\mbox{for PPN})
\label{eq: Effective magnetic moment of PPN}
%\nonumber\\
\end{eqnarray}
We note that the parameter of the theoretical model of gravity $\gamma^*$ explicitly appears in the gravitationally induced correction term.
This implies that the precise determination of the magnetic moment in a specific local circumstance may determine the model parameter $\gamma^*$.

\section{Penning Trap Experiment to Measure Electron g-2 in The Earth's Gravity}
%
%%%%%%%%%%%%%
The g-factor of leptons exactly equals to 2 according to the Dirac theory; 
(${\rm g}_l$=2, for $l\!=\!{\rm e},\mu,\tau$), which is one of important consequences of the Dirac equation for free fermions with the electric charge of $e$ and the spin of $1/2$ minimally coupling to electromagnetic field.
However, the real value of ${\rm g}_l$ deviates from 2 with the fraction of about $0.002$ as a result of the quantum radiative corrections.
The deviation is referred to as the anomalous magnetic moment defined as
\begin{equation}
  a_{l} 
    \equiv   \frac{{\rm g}_{l}}{2} -1 
     =       \frac{\mu_{\rm m}^l}{\mu_{\rm B}^l} -1    \hspace{3mm}  ( = 0.001... )
,
\label{eq: a=g-2/2}
\end{equation}
where $\mu_{\rm m}^l$ is the magnetic moment and ${\mu_{\rm B}^l}$ is the Bohr magneton. 
The Bohr magneton is defined as ${\mu_{\rm B}^l}$=$(e/m_l)\,S$ for the lepton mass $m_l$ and the spin $S$.
For electrons, the comparison between the theoretical value, which was obtained with higher orders of quantum radiative corrections based on the standard model of elementary particles (SM)~\cite{Aoyama-2015,CODATA-2014}, and the experimental value, which was obtained in the precision measurement~\cite{electron-g-2-experiment,PDG-2016,CODATA-2014}, shows the extremely precise agreement up to the 12th digit
\begin{eqnarray}
  a_{{\rm e (EXP)}} - a_{{\rm e (SM)}} = -0.91 \,\, (0.82) \times\! 10^{-12}
\label{eq: g-2 agreement of electrons}
\end{eqnarray}
as shown in Table~\ref{tab: g-2 values for electron}, which clearly validates the accuracy of the quantum electrodynamics (QED) employed in the standard model of elementary particles~\cite{Aoyama-2015,PDG-2016,CODATA-2014}.
\begin{table}[!h]
\caption{\label{tab: g-2 values for electron} 
Comparison of the theoretical value and the experimental value of the anomalous magnetic moment of electron.
}
\centering
\begin{tabular}{|c||l|c|}
\hline
 & & \\
    & \,\,\,\,\, anomalous magnetic moment   & {\scriptsize Ref.}  \\
    &  \hspace{10mm} ($a_e\equiv {\rm g}_e/2-1$)          &            \\
 & & \\
\hline
% & & \\
 & & \\
{\scriptsize Experiment} 
      &   $a_{{\rm e (EXP)}}$      \!\!=\! 1\,159\,652\,180.73 \, (0.28)  $\times10^{-12}$
      &   \cite{electron-g-2-experiment,PDG-2016,CODATA-2014} \\
{\scriptsize Theory} 
      &   $a_{{\rm e (SM)}}$\,\,\, \!\!=\! 1\,159\,652\,181.643   (0.77)  $\times10^{-12}$
      &   \cite{Aoyama-2015,CODATA-2014} \\
 & & \\
{\scriptsize Difference}
      &   $a_{{\rm e (EXP)}} \!-\! a_{{\rm e (SM)}}$   $= -0.91  $ \,\,\,\,(0.82) $\times10^{-12}$ 
      &    \\
{\scriptsize (Deviation)}
      &    \hspace{20mm} (1.1 $\sigma$) 
      &    \\
 & & \\
\hline
\end{tabular}
\end{table}

\subsection{Interpretation of the experimental result of electron g-2}
Below, we examine the interpretation of the experimental result of the Penning trap experiment~\cite{electron-g-2-experiment,CODATA-2014}.

The cyclotron frequency $\Omega_{\rm c}$ and the spin precession frequency $\Omega_{\rm s}$ of an electron moving in an uniform magnetic field are measured in the Penning trap experiment~\cite{electron-g-2-experiment,CODATA-2014}.
The anomalous magnetic moment of electrons $a_{\rm e}$ is regarded equal to the ratio of the anomalous spin precession frequency $\Omega_a\!\equiv\!\Omega_s\!-\!\Omega_c$ to the cyclotron frequency $\Omega_{\rm c}$ as
\begin{eqnarray}
	a_{\rm e}^{\rm conv}\!{}_{({\rm EXP})} = \frac{\Omega_a}{\Omega_c}= \frac{\Omega_s}{\Omega_c}-1
	.
	\label{eq: a_e}
\end{eqnarray}

The ${a_{\rm e}^{\rm conv}}_{\rm (EXP)}$ has been treated as the electron anomalous magnetic moment in the conventional analysis without the consideration of gravitational effect.
However, this experiment was carried out in the Earth's gravitational field.
Thus the experimentally measured frequencies in Eq.~(\ref{eq: a_e}) are the effective frequencies in a curved spacetime.
Hereafter, we derive the effective values of the cyclotron frequency $\Omega_{\rm c}^{\rm eff}$ and the spin precession frequency $\Omega_{\rm s}^{\rm eff}$ in a curved spacetime.

The general relativity requires that the translational motion of a free particle with the electric charge of $e$ is described by the covariant equation of Eq.(\ref{eq: translational motion on curved space}).
Assuming the Earth's gravitational field can be described with the Schwarzschild metric shown in Eq.~(\ref{eq: Schwarzschild-metric}), 
the equation of translational motion of a free electron can be written as Eq.(\ref{eq: translational motion on curved space-2}). 
In general, the solution of this equation gives a spiral motion which is the combination of the translational acceleration and the rotational motion.
The cyclotron frequency is the measure of the rotational motion.
In other words, 
the cyclotron frequency corresponds to the frequency of the rotational motion of the constant-norm vector parallel to the velocity.
We define the unit vector parallel to the three-dimensional velocity $\bm{\beta}$ as
\begin{eqnarray}
  {\bm \beta}  \equiv  \beta\,\,{\bm {\hat \beta}}
               \,\,\, , \,\,\,\,\,\,\,\,\, 
  {\bm {\hat \beta}} \cdot {\bm {\hat \beta}} = \frac{{\bm \beta}\cdot{\bm \beta}}{\beta^2} = 1
.
\end{eqnarray}
Here we use $\bm{\beta}\!\cdot\!\nabla\phi=0$ and assume the effects of the gravitational gradient is negligibly small, 
the rotational component of the translational motion is given as
\begin{eqnarray}
	\frac{d {\bm {\hat \beta}} }{dt}
	&=& 
		- \Big(\! 1\!+\!(2\gamma^2\!+\!1)\epsilon^2\phi\! \Big)\,\,
		\frac{e}{m} \Big(\frac{{\bm B}}{\gamma} \!-\!  \frac{\gamma}{\gamma^2\!-\!1} ({\bm \beta}\!\times\!{\bm E}) \Big)
		\!\times\! {\bm {\hat \beta}}
	\nonumber\\
	&=&
		{\bm \Omega_c}^{\rm eff} \,\times \, {\bm {\hat \beta}}
.
\end{eqnarray}
Using the cyclotron frequency in the flat spacetime
$
{\bm \Omega}_c\!=
    -e/m
     \Big(\! 
            {\bm B}/\gamma -  \gamma/(\gamma^2\!-\!1)\, ({\bm \beta}\!\times\!{\bm E}) 
     \!\Big)
,
$
the effective cyclotron frequency can be written as
\begin{eqnarray}
{\bm \Omega}_{\rm c}^{\rm eff}
	&=& 
	\,\,\,\, \Big(\! 1\!+\!(2\gamma^2\!+\!1)\epsilon^2\phi\! \Big)\,\, {\bm \Omega}_c
	\,\,\, + O(\epsilon^4) 
	,
\label{eq: Omega_c by GR}
\end{eqnarray}
which represents the cyclotron frequency in the Earth's gravitational field based on the general relativity.

In the measurement of the electron ${\rm g}_e\!-\!2$ experiment (i.e., Penning trap experiment), the electron velocity is sufficiently small as $\beta\ll1$ and the effective value of the electron's cyclotron frequency is given as
\begin{eqnarray}
  {\bm \Omega}_{\rm c}^{\rm eff} =  (1\!+\!3\epsilon^2\phi) \,\, {\bm \Omega}_c\,\,.
\end{eqnarray}
We can interpret that the ${\bm \Omega}_{\rm c}^{\rm eff}$ is used as the experimentally measured cyclotron frequency to be observed with the ground-based instrument.

Here we analyze the behavior of an electron spin in the Earth's gravitational field.
The time evolution of an electron spin in the flat spacetime is described by the equation
\begin{eqnarray}
     \frac{d {\bm S}}{d t} 
  &=& {\bm \mu}_{\rm m}\!\times\!{\bm B} 
   = {\bm \Omega}_{\rm s}\!\times\!{\bm S} 
   ,
\label{eq: spin motion at rest-frame in classical}
\end{eqnarray}
where $\bm{\Omega}_{\rm s}=-({\rm g}_{\rm e}/2)(e/m)\,\bm{B}$.
Rewriting this equation with the four vector, we obtain
\begin{eqnarray}
  \frac{d S^\mu}{d \tau} 
 &=&
  \frac{{\rm g}_e}{2}\frac{e}{m} 
    \Bigg(\!
          F^{\mu\nu} S_\nu 
         + \frac{1}{c^2} u^\mu (S_\lambda F^{\lambda \nu} u_\nu ) \!
    \Bigg) 
         - \frac{1}{c^2} u^\mu (S_\lambda \frac{d u^\lambda}{d \tau} ) 
,
\label{eq: BMT equation}
\end{eqnarray}
which is known as the BMT equation~\cite{BMT-equation}.
In the same manner as the Eq.(\ref{eq: translational motion on curved space}), employing the covariant derivative into the BMT equation, we obtain the covariant equation of spin motion in the general relativity as
\begin{eqnarray}
  \frac{D S^\mu}{d \tau} 
 &=&
  \frac{{\rm g}_e}{2}\frac{e}{m} 
    \Bigg(\!
          F^{\mu\nu} S_\nu 
         + \frac{1}{c^2} u^\mu (S_\lambda F^{\lambda \nu} u_\nu ) \!
    \Bigg) 
         - \frac{1}{c^2} u^\mu (S_\lambda \frac{D u^\lambda}{d \tau} ) 
.
\label{eq: BMT+GR equation}
\end{eqnarray}
Here we substitute the Schwarzschild metric defined by Eq.~(\ref{eq: Schwarzschild-metric}) into the covariant BMT equation.
In the same way as we have derived the cyclotron frequency based on the equation of translational motion of electrons, 
we consider the case where the gradient of the gravitational potential $\nabla\phi$ is negligibly small such as the case where the motion is limited on a horizontal plane.
We define the four spin vector as $s^{\mu}=(S^0,\bm{S})$ and the spatial component of the generalized BMT equation Eq.~(\ref{eq: BMT+GR equation}) up to the post-Newtonian order $O(\epsilon^2)$ can be written as
\begin{eqnarray}
        \frac{d {\bm S}}{d t} 
     = 
        {\bm S} \times \frac{\Big(1+\epsilon^2\phi(2\gamma^2+1)\Big)}{\gamma}\frac{\rm g}{2}\frac{e}{m}{\bm B}
.
\end{eqnarray}
We can take $\beta\ll1$ in the measurement of electron ${\rm g}_e\!-\!2$ ({\it i.e.,} Penning trap experiment), which implies that the effective value of the electron spin precession frequency to be observed with the ground-based instrument is given as
\begin{eqnarray}
{\bm \Omega}_{\rm s}^{\rm eff}
   &=& 
      ( 1+3\epsilon^2\phi )\,\, {\bm \Omega}_s
.
\end{eqnarray}

The experimental value of the anomalous magnetic moment obtained in the Penning trap experiment, denoted as $a_{\rm e(EXP)}^{\rm eff}$ below, should be compared with the ratio of the kinematically calculated effective values $\Omega_{\rm c}^{\rm eff}$ and $\Omega_{\rm s}^{\rm eff}$.
Substituting the effective values $\Omega_{\rm c}^{\rm eff}$ and $\Omega_{\rm s}^{\rm eff}$ into Eq.~(\ref{eq: a_e}), we obtain
\begin{eqnarray}
       a_e^{\rm eff} {}_{({\rm EXP})}
 &=&    
       \frac{\Omega_s^{\rm eff}}{\Omega_c^{\rm eff}}-1
  =    \frac{(1\!+\!3\epsilon^2\phi)\,\,{\Omega}_s}{(1\!+\!3\epsilon^2\phi)\,\,{\Omega}_c}-1
       \nonumber\\
 &=&   \frac{\Omega_s}{\Omega_c}-1 
       \nonumber\\
 &=&   a_e^{\rm conv}\!{}_{({\rm EXP})} 
.
\end{eqnarray}
Although both effective values $\Omega_{\rm c}^{\rm eff}$ and $\Omega_{\rm s}^{\rm eff}$ are dependent of the Earth's gravitational field,
gravitational effects are canceled in their ratio and the $a_{\rm e(EXP)}^{\rm eff}$ coincides with $a_{\rm e(EXP)}^{\rm conv}$.
It is reasonable to regard that the high precision agreement between the experimental and theoretical values is confirming the cancellation of gravitational effects.
It also suggests that, if we take the ratio of the conventionally defined $a_e$ in Eq.~(\ref{eq: a=g-2/2}) to the Bohr magneton, the ``true'' anomalous magnetic moment $a_e^{\rm eff}\!{}_{({\rm EXP})}$ deviates from the ``conventional'' value $a_e^{\rm conv}\!{}_{({\rm EXP})}$ as
\begin{eqnarray}
 a_e^{\rm eff}\!{}_{({\rm EXP})} 
             &\equiv& \frac{{\rm g}_e^{\rm eff}}{2} - 1 
              =  \frac{\mu_{\rm m}^{\rm eff}}{\mu_{\rm B}}-1 
              =  \frac{\Omega_s^{\rm eff}}{\Omega_c}-1
                 \nonumber\\
             &=& (1\!+\!3\epsilon^2\phi)\,\frac{\Omega_s}{\Omega_c}-1
                 \nonumber\\
             &\ne& \frac{\Omega_s}{\Omega_c}-1 
             =  a_e^{\rm conv}\!{}_{({\rm EXP})} 
.
\end{eqnarray}
Since the Earth's gravity induces an anomaly even for ${\rm g_e}$=2 additionally to radiative corrections,
the comparison between experimental value measured in the curved spacetime and the theoretical value calculated in the flat spacetime results in the difference of
\begin{eqnarray}
	\left\vert a_e^{\rm eff}\!{}_{({\rm EXP})} -  a_e{}_{({\rm SM})} \right\vert
	&\simeq&   3\epsilon^2|\phi| 
	\sim 2.1 \times 10^{-9}
.
\end{eqnarray}
Here we introduce the redefinition of the effective Bohr magneton including the general relativistic effects of the Earth's gravitational field as
\begin{eqnarray}
\mu_{\rm B}^{\rm eff} 
    &\equiv& 
        \mu_{\rm B}^{\rm eff}(\phi) 
      \simeq
        (1\!+\!3\epsilon^2\phi)\,\mu_{\rm B} 
.
\end{eqnarray}
In this case, we can treat the $a_{\rm e(EXP)}^{\rm eff}$ as the invariant quantity independent of the gravitational field if we cancel the general relativistic effects by redefining the anomalous magnetic moment using the ratio of the effective magnetic moment to the effective Bohr magneton as
\begin{eqnarray}
 a_e^{\rm eff}{}_{({\rm EXP})} 
	&\equiv& 
		\frac{\mu_{\rm m}^{\rm eff}}{\mu_{\rm B}^{\rm eff}}-1 
	=  \frac{\Omega_s^{\rm eff}}{\Omega_c^{\rm eff}}-1
		\nonumber\\
	&=&  a_e^{\rm conv}{}_{({\rm EXP})} 
,
\end{eqnarray}
instead of its ratio to the Bohr magneton in the flat spacetime.
The redefinition restores the validity of the test of the standard model of elementary particles via the comparison of the theoretical value $a_{\rm e(SM)}$ and the experimental value $a_e^{\rm eff}\!{}_{({\rm EXP})}$.

\section{Storage Ring Experiment to Measure Muon g-2 in The Earth's Gravity}
In the storage ring experiments for ${\rm g}_\mu\!-\!2$ such as CERN \cite{muon-g-factor-CERN} and BNL E821 \cite{muon-g-factor-BNL}, the muon anomalous magnetic moment $a_{\mu ({\rm EXP})}$ was determined from the relation
\begin{eqnarray}
a_{\mu ({\rm EXP})}   
   &\equiv& \frac{R}{\lambda - R}
,
  \label{eq: a=R/lambda-R}
\end{eqnarray}
where $R \equiv {\Omega_{\rm a}}/{\Omega_{\rm p}}$ is the ratio of the anomalous spin precession frequency $\Omega_{\rm a}$ of muons to the proton spin precession frequency $\Omega_{\rm p}$ measured using the nuclear magnetic resonance and $\lambda\equiv {\mu_\mu }/{\mu_p}$ is the ratio of the magnetic moment of muons to that of protons.
This can be understood as that the $a_{\mu ({\rm EXP})}$ is determined from the relation
\begin{eqnarray}
	a_{\mu ({\rm EXP})}   
	&=&      \frac{\Omega_a}{\Omega_L - \Omega_a}
	,
	\label{eq: a=Omega_a/Omega_L-Omega_a}
\end{eqnarray}
where 
$\Omega_{\rm L}$ the Larmor precession frequency of muons.
This interpretation can be obtained using the classical kinematical framework and it is the basis of the design of the experiment to measure the muon ${\rm g}_\mu\!-\!2$~\cite{muon-g-factor-BNL,muon-g-factor-CERN,PDG-2016,CODATA-2014}.
We reconsider this interpretation including the spacetime curvature in the Earth's gravitational field on the basis of general relativity.

\subsection{Spin Motion in Flat Spacetime}
Here we overview the procedure to evaluate the experimental value of the anomalous magnetic moment using the equation of motion of spin in the flat spacetime (e.g. See Chap.11 in Ref.~\cite{jackson-1998}).
The BMT equation Eq.~(\ref{eq: BMT equation}) described in the laboratory frame (inertial frame) can be decomposed into a set of equations for time and spatial components as
\begin{eqnarray}
	\frac{d S^0}{d \tau}     
	&=&
		F_0     
		+
		\,\,\gamma^2 \, ({\bm S}\!\cdot\!\frac{d{\bm \beta}}{d\tau}) 
		\nonumber\\
	\frac{d {\bm S}}{d \tau} 
		&=&
		{\bm F} 
		+ 
		\,\,\gamma^2 {\bm \beta}\, ({\bm S}\!\cdot\!\frac{d{\bm \beta}}{d\tau}) 
.
\label{eq: BMT in inertial-flame}
\end{eqnarray}
The three-dimensional spin vector $\bm{S}$ in the inertial frame and the three-dimensional spin vector $\bm{s}$ are related through the Lorentz transformation
\begin{eqnarray}
	{\bm s} &=&   {\bm S} - \frac{\gamma}{\gamma+1}({\bm \beta}\cdot{\bm S})   {\bm \beta}   \nonumber\\
	&=&   {\bm S} - \frac{\gamma}{\gamma+1} S^0 {\bm \beta}
.
\label{eq: Lorentz-transform-of-spin}
\end{eqnarray}
The equation of motion of the spin in the rest frame $\bm{s}$ can be derived as
\begin{eqnarray}
	\frac{d{\bm s}}{d\tau}
	&=&   \Big(
		{\bm F} 
		- \frac{\gamma}{\gamma+1} F_0 \, {\bm \beta} 
		\Big)
		+  \frac{\gamma^2}{\gamma+1}\,\,{\bm s} 
		\!\times\! 
		({\bm \beta}\!\times\!{\frac{d{\bm \beta}}{d\tau}})  \nonumber\\
	&=&   {\bm F}' 
		+  \frac{\gamma^2}{\gamma+1}\,\,{\bm s} 
		\!\times\! 
		({\bm \beta}\!\times\!{\frac{d{\bm \beta}}{d\tau}})  
.
\end{eqnarray}
Rewriting the proper time $\tau$ with the time in the inertial frame $t$, we can reproduce the familiar equation
\begin{eqnarray}
	\frac{d{\bm s}}{dt}
	&=&   
		\frac{\bm F'}{\gamma} 
		+  \frac{\gamma^2}{\gamma+1}\,\,{\bm s} 
		\!\times\! 
		({\bm \beta}\!\times\!{\frac{d{\bm \beta}}{dt}})  \nonumber\\
	&=&  
		\frac{{\rm g}_\mu}{2} \frac{e}{\gamma\,m}\,\, {\bm s}\!\times\!{\bm B'}      
		+  \frac{\gamma^2}{\gamma+1}\,\,{\bm s} 
		\!\times\! 
		({\bm \beta}\!\times\!{\frac{d{\bm \beta}}{dt}})
.
\label{eq: spin motion at rest-frame in special relativity}
\end{eqnarray}
We note that the last term in the equation of motion of spin in the rest frame (Eq.~(\ref{eq: spin motion at rest-frame in special relativity})) represents the relativistic term known as the Thomas precession which does not appear in the non-relativistic equation of motion of spin (Eq.~(\ref{eq: spin motion at rest-frame in classical})).
The electric and magnetic fields in the inertial frame $\bm{E}$ and $\bm{B}$ are related with those in the rest frame $\bm{E}'$ and $\bm{B}'$ through the Lorentz transformation as
\begin{eqnarray}
	{\bm E'} &=&   \gamma ({\bm E}+{\bm \beta}\!\times\!{\bm B}) 
		- \frac{\gamma^2}{\gamma+1}\,\,({\bm \beta}\!\cdot\!{\bm E})\,{\bm \beta}\, 
		\nonumber\\
	{\bm B'} &=&   \gamma ({\bm B}-{\bm \beta}\!\times\!{\bm E}) 
		- \frac{\gamma^2}{\gamma+1}\,\,({\bm \beta}\!\cdot\!{\bm B})\,{\bm \beta}\,
.
\label{eq: Lorentz transform of E-B}
\end{eqnarray}
The equation of motion of spin in the rest frame can be rewritten with the quantities in the inertial frame by substituting Eq.~(\ref{eq: translational motion on flat space-2}) and Eq.~(\ref{eq: Lorentz transform of E-B}), which results in
\begin{eqnarray}
	\frac{d{\bm s}}{dt}
	&=&    \frac{\rm g}{2} \frac{e}{m}\,\, {\bm s}
		\!\times\!
		\Big(
			{\bm B}-{\bm \beta}\!\times\!{\bm E} 
			- \frac{\gamma}{\gamma+1}\,\,({\bm \beta}\!\cdot\!{\bm B})\,{\bm \beta} 
		\Big)
		\nonumber\\
	& &     
                \hspace{7mm}
		+   \frac{\gamma^2}{\gamma\!+\!1}\,\,{\bm s} 
		\!\times\! 
		\Big(
			{\bm \beta}
			\!\times\!
			\frac{e}{m\gamma}\,\, 
			\big(
				{\bm E}
				+ {\bm \beta}\!\times\!{\bm B}
				- ({\bm \beta}\!\cdot\!{\bm E})\,{\bm \beta}
			\big)
		\Big)
		\nonumber\\
	&=&    {\bm s}
		\!\times\!
		\frac{e}{m}
		\Bigg(
			(\frac{{\rm g}}{2}\!-\!1\!+\!\frac{1}{\gamma}){\bm B}
			-  (\frac{{\rm g}}{2}\!-\!\frac{\gamma}{\gamma+1}){\bm \beta}\!\times\!{\bm E} 
			-  (\frac{{\rm g}}{2}\!-\!1)\frac{\gamma}{\gamma+1}\,({\bm \beta}\!\cdot\!{\bm B}) \,{\bm \beta}
		\Bigg)
		\nonumber\\
	&=&    {\bm \Omega}_{\rm s} \times {\bm s}
.
\label{eq: spin motion at inertial-frame in special relativity}
\end{eqnarray}
By using the spin precession frequency ${\bm \Omega}_{\rm s}$, the anomalous spin precession frequency ${\bm \Omega}_{\rm a}$ can be written as
\begin{eqnarray}
	{\bm \Omega}_{\rm a}
	&=&
		{\bm \Omega}_{\rm s}
		- {\bm \Omega}_{\rm c}
		\nonumber\\
	&=&
		- \frac{e}{m}
		\Bigg(
		a_\mu\,{\bm B}
		-  (a_\mu\!-\!\frac{1}{\gamma^2\!-\!1})\,{\bm \beta}\!\times\!{\bm E} 
		-  a_\mu \frac{\gamma}{\gamma\!+\!1} \,({\bm \beta}\!\cdot\!{\bm B}) \,{\bm \beta}
		\Bigg),
\label{eq: Omega_a by SR}
\end{eqnarray}
with $a_\mu\equiv {\rm g}_\mu/2\!-\!1$.
We assume that the motion is confined in the horizontal plane and the magnetic field is applied vertically.
In addition, we assume that the electric term (the second term) can be canceled as $\big( a_\mu \!-\!1/(\gamma^2\!-\!1) \big) \!=\!0$ by selecting the muon momentum of $\gamma\simeq$ 29.3, which is referred to as the magic momentum.
The anomalous magnetic moment of muon in the flat spacetime, which is defined as
$a_{\mu {\rm (EXP)}} \equiv -{  \Omega_a  }/{ (e\,B/m) } $, can be written as
\begin{eqnarray}
	a_{\mu ({\rm EXP})}
		&\equiv& - \frac{ { \Omega}_a  }{ (\frac{e { B}}{m}) } 
		=         \frac{ { \Omega}_a  }{ { \Omega}_L - { \Omega}_a } 
		=         \frac{ { \Omega_a}/{\Omega_p}  }{ { \Omega_L}/{\Omega_p} - {\Omega_a}/{\Omega_p} } 
		\nonumber\\
	&=&       \frac{R}{\lambda - R}
,
\label{eq: a_mu_exp^conv}
\end{eqnarray}
by rewriting the magnetic field $\bm{B}$ using the Larmor precession frequency $\Omega_{\rm L}=-({\rm g}_\mu/2)(e/m)\,{B}=-(1+a_\mu)(e/m){B}$.
The Eq.~(\ref{eq: a_mu_exp^conv}) represents the experimental value in the conventional interpretation.
The experimental value of the anomalous magnetic moment $a_{\mu ({\rm EXP})}$ using Eq.~(\ref{eq: Omega_a by SR}) and Eq.~(\ref{eq: a_mu_exp^conv}) are evaluated within the classical kinematical framework.
The relations were applied to achieve the high precision by employing the magic momentum to suppress the contribution of electric fields.

\subsection{Spin Motion in Curved Spacetime}
In this section, we apply the procedure of the evaluation of the experimental value of the anomalous magnetic moment in the flat spacetime to that in the curved spacetime.

The BMT equation in the curved spacetime in the inertial frame (Eq.~(\ref{eq: BMT+GR equation})) can be decomposed into the time and spatial components as
\begin{eqnarray}
&&  \hspace{0mm} \frac{d S^0}{d \tau}     
   \!+\! \epsilon^2 ( S^0 {\bm u} \!+\! u^0 {\bm S} )\!\cdot\!\nabla\phi
                    \nonumber\\
&& \hspace{7mm} =   
                    F_0     
                  + ( 1\!-\!4\gamma^2\epsilon^2\phi ) 
                    \,\,\gamma^2 \, ({\bm S}\!\cdot\!\frac{d{\bm \beta}}{d\tau}) 
                  \,\,\, + O(\epsilon^4)
                    \nonumber\\
                    \nonumber\\
&&  \hspace{0mm}  \frac{d {\bm S}}{d \tau} 
   \!+\! \epsilon \,\, u^0 S^0 \, \nabla\phi 
   \!+\! \epsilon^2 \Big(
                        ({\bm u}\!\cdot\!{\bm S})\nabla\phi \!-\! ({\bm S}\!\cdot\!\nabla\phi){\bm u} \!-\! ({\bm u}\!\cdot\!\nabla\phi){\bm S}
                    \Big) \nonumber\\
&& \hspace{7mm} =
                   {\bm F} 
                  + ( 1\!-\!4\gamma^2\epsilon^2\phi )
                    \,\,\gamma^2 {\bm \beta}\, ({\bm S}\!\cdot\!\frac{d{\bm \beta}}{d\tau}) 
                  \,\,\,+ O(\epsilon^4)
.
\label{eq: BMT+GR in inertial-flame}
\end{eqnarray}
Considering the contribution arising from the gradient of the gravitational potential $\nabla\phi$ is sufficiently small to be ignored in the post-Newtonian order (See Appendix~\ref{app: typical scale of nabla-phi}), 
Eq.~(\ref{eq: BMT+GR in inertial-flame}) can be simplified as
\begin{eqnarray}
	\frac{d S^0}{d \tau}     
	&=&
		F_0     
		\!+\! ( 1\!-\!4\gamma^2\epsilon^2\phi )
		\,\,\gamma^2 \, ({\bm S}\!\cdot\!\frac{d{\bm \beta}}{d\tau}) 
		\,\,+ O(\epsilon^4)
		\nonumber\\
	\frac{d {\bm S}}{d \tau} 
	&=&
		{\bm F} 
		\!+\! ( 1\!-\!4\gamma^2\epsilon^2\phi )
		\,\,\gamma^2 {\bm \beta}\, ({\bm S}\!\cdot\!\frac{d{\bm \beta}}{d\tau}) 
		\,\,+ O(\epsilon^4)
.
\label{eq: BMT+GR in inertial-flame-2}
\end{eqnarray}
Following the procedure in the flat spacetime, we derive the relation between the spins in the inertial frame and the rest frame using the Lorentz transformation.
The Lorentz transformation can be naturally defined in the flat spacetime since the Lorentz invariance is satisfied at any point.
However, the Lorentz transformation cannot be defined in the general coordinate system (curved spacetime), which disturbs the applicability of the same procedure.
We avoid the difficulty by employing the local inertial frame to satisfy the local Lorentz invariance (See Appendix~\ref{app:LLI}).
The four vector in the general coordinate system $x^\mu\!=\!(x^0, {\bm x})$ can be related with the four vector in the local inertial frame $x^{(a)}\!=\!(\tilde{x}^0, \tilde{\bm x})$, using the tetrad $e^{(a)}_\mu$, as
\begin{eqnarray}
	x^{(a)}    &\equiv&   e^{(a)}_{\mu} \,x^{\mu} = (\tilde{x}^{0}, \tilde{{\bm x}}) 
.
\end{eqnarray}
Using the explicit expression of the tetrad up to the post-Newtonian order $O(\epsilon^2)$ in the Schwarzschild metric (shown in Eq.(\ref{app_eq: tetrad})), the transformation from the general coordinate system into the local inertial frame can be simplified as
\begin{eqnarray}
	\tilde{S}^0       &=&  (1+\epsilon^2\phi) \, {S}^0     \nonumber\\
	\tilde{\bm S}     &=&  (1-\epsilon^2\phi) \, {\bm S}   \nonumber\\
	\tilde{\bm \beta} &=&  (1-2\epsilon^2\phi) \,{{\bm \beta }}  \nonumber\\
	\tilde{ t}        &=&  (1+\epsilon^2\phi) \,\,t    \nonumber\\
	\tilde{ \gamma }  &=&  \Big( 1-2\epsilon^2\phi\, (\gamma^2-1) \Big) \,{{    \gamma}}   \nonumber\\
	\tilde{\bm E}     &=&  {\bm E}	\nonumber\\ 
	\tilde{\bm B}     &=&  (1+2\epsilon^2\phi) {\bm B} \nonumber\\ 
	& & ... \,\,\,\, {\rm etc}
\end{eqnarray}
We rewrite the BMT equation in the curved spacetime (Eq.~(\ref{eq: BMT+GR in inertial-flame-2})) using the local inertial frame and obtain
\begin{eqnarray}
&& 
	\frac{d \tilde{S}^0}{d \tau}     
	=
	\tilde{F}_0     
	+
	\,\tilde{\gamma}^2 \,(\tilde{\bm S}\!\cdot\!\frac{d \tilde{\bm \beta}}{d\tau}) 
	+ O(\epsilon^4)
	\nonumber\\
&& 
	\frac{d \tilde{\bm S}}{d \tau} 
	=
	\tilde{\bm F} 
	+ \tilde{\gamma}^2 
	\tilde{\bm \beta}\, ( \tilde{\bm S}\!\cdot\!\frac{d \tilde{\bm \beta}}{d\tau}) 
	+ O(\epsilon^4)
,
\label{eq: BMT+GR in inertial-flame-3}
\end{eqnarray}
which is an analogue to the relation in the flat spacetime.
The local Lorentz transformation is obtained as 
\begin{eqnarray}
\tilde{\bm s}
	&=&  \tilde{\bm S} - \frac{\tilde{\gamma}}{\tilde{\gamma}+1} (\tilde{\bm \beta}\cdot \tilde{\bm S}) \tilde{\bm \beta}   	\nonumber\\
	&=&  \tilde{\bm S} - \frac{\tilde{\gamma}}{\tilde{\gamma}+1} \tilde{S}^0 \tilde{\bm \beta}   
.
\label{eq: LLT}
\end{eqnarray}
In the same manner as the flat spacetime, using the BMT equation in the curved spacetime (Eq.~(\ref{eq: BMT+GR in inertial-flame-3})) and the local Lorentz transformation (Eq.~(\ref{eq: LLT})), the equation of motion of spin in the rest frame can be written as
\begin{eqnarray}
\frac{d\tilde{\bm s}}{d\tau}
	&=&   \Big(
		\tilde{\bm F} 
		- \frac{\tilde{\gamma}}{\tilde{\gamma}+1} \tilde{F}_0 \, \tilde{\bm \beta} 
	\Big)
	+  \frac{\tilde{\gamma}^2}{\tilde{\gamma}+1}\,\,\tilde{\bm s} 
		\!\times\! 
		(\tilde{\bm \beta}\!\times\!{\frac{d \tilde{\bm \beta}}{d\tau}})  \nonumber\\
	&=&   \tilde{\bm F}' 
	+  \frac{\tilde{\gamma}^2}{\tilde{\gamma}+1}\,\,\tilde{\bm s} 
		\!\times\! 
		(\tilde{\bm \beta}\!\times\!{\frac{d \tilde{\bm \beta}}{d\tau}})
,
\end{eqnarray}
which is similar to the case of the flat spacetime.
This relation gives the relation similar to the case of the flat spacetime
\begin{eqnarray}
 \frac{d \tilde{\bm s}}{d \tilde{t}}
    &=&   
         \frac{\tilde{\bm F}'}{\tilde{\gamma}} 
      +  \frac{\tilde{\gamma}^2}{\tilde{\gamma}+1}\,\,\tilde{\bm s} 
          \!\times\! 
         (\tilde{\bm \beta}\!\times\!{\frac{d \tilde{\bm \beta}}{d \tilde{t}}})  \nonumber\\
    &=&  
         \frac{{\rm g}_\mu}{2} \frac{e}{\tilde{\gamma}\,m}\,\, \tilde{\bm s}\!\times\!\tilde{\bm B'}      
      +  \frac{\tilde{\gamma}^2}{\tilde{\gamma}+1}\,\,\tilde{\bm s} 
          \!\times\! 
         (\tilde{\bm \beta}\!\times\!{\frac{d \tilde{\bm \beta}}{d \tilde{t}}})
,
\label{eq: spin motion at rest-frame in curved spacetime}
\end{eqnarray}
by replacing the proper time $\tau$ with the time in the inertial frame $\tilde{t}$.
Here we notice that the electric and magnetic fields in the rest frame $\tilde{\bm E'}$ and $\tilde{\bm B'}$ is given through a local Lorentz transformation, using the electric and the magnetic fields in the inertial frame $\tilde{\bm E}$ and $\tilde{\bm B}$, as
\begin{eqnarray}
\tilde{\bm E'} &=&   \tilde{\gamma} (\tilde{\bm E}+\tilde{\bm \beta}\!\times\!\tilde{\bm B}) 
                    - \frac{\tilde{\gamma}^2}{\tilde{\gamma}+1}\,\,(\tilde{\bm \beta}\!\cdot\!\tilde{\bm E})\,\tilde{\bm \beta}
                      \nonumber\\
\tilde{\bm B'} &=&   \tilde{\gamma} (\tilde{\bm B}-\tilde{\bm \beta}\!\times\!\tilde{\bm E}) 
                    - \frac{\tilde{\gamma}^2}{\tilde{\gamma}+1}\,\,(\tilde{\bm \beta}\!\cdot\!\tilde{\bm B})\,\tilde{\bm \beta} 
.
\label{eq: LLT of E-B}
\end{eqnarray}
Thus, the equation of motion of spin in the rest frame $\tilde{\bm s}$ can be written using the electric and the magnetic fields in the inertial frame $\tilde{\bm E}$ and $\tilde{\bm B}$ as
\begin{eqnarray}
\frac{d \tilde{\bm s}}{d \tilde{t}}
	\!\!&=&\!\!    \frac{\rm g_\mu}{2} \frac{e}{m}\,\, \tilde{\bm s}
	\!\times\!
	\Big(
		\tilde{\bm B}-\tilde{\bm \beta}\!\times\!\tilde{\bm E} 
		- \frac{\tilde{\gamma}}{\tilde{\gamma}+1}\,\,(\tilde{\bm \beta}\!\cdot\!\tilde{\bm B})\, \tilde{\bm \beta}
	\Big)
	\nonumber\\
	& &
        \hspace{7mm}
	+   \frac{\tilde{\gamma}^2}{\tilde{\gamma}\!+\!1}\,\,\tilde{\bm s} 
	\!\times\! 
	\Big(
		\tilde{\bm \beta}
		\!\times\!
		\frac{e}{m\tilde{\gamma}}\,\, 
		\big(
			\tilde{\bm E} 
			+ \tilde{\bm \beta}\!\times\!\tilde{\bm B} 
			- (\tilde{\bm \beta}\!\cdot\!\tilde{\bm E})\,\tilde{\bm \beta}
		\big)
	\Big)
	\nonumber\\
	&=&    \tilde{\bm s}
		\!\times\!
		\frac{e}{m}
		\Bigg(
			(\frac{{\rm g_\mu}}{2}\!-\!1\!+\!\frac{1}{\tilde{\gamma}})\tilde{\bm B}
			-  (\frac{{\rm g_\mu}}{2}\!-\!\frac{\tilde{\gamma}}{\tilde{\gamma}+1})\tilde{\bm \beta}\!\times\!\tilde{\bm E} 
			-  (\frac{{\rm g_\mu}}{2}\!-\!1)\frac{\tilde{\gamma}}{\tilde{\gamma}+1} (\tilde{\bm \beta}\!\cdot\!\tilde{\bm B})\,\tilde{\bm \beta} 
		\Bigg)
.
\label{eq: spin motion at inertial-frame in curved spacetime}
\end{eqnarray}

The Eq.(\ref{eq: spin motion at inertial-frame in curved spacetime}) has the same form as the equation of spin motion Eq.(\ref{eq: spin motion at inertial-frame in special relativity}).
We convert the electric and magnetic fields $\tilde{\bm E}$ and $\tilde{\bm B}$ in the local inertial frame to those in the general coordinate system to consider the experimental values with the ground-based instrument.

The equation of motion of spin ${\bm s}$ in the rest frame represented in the general coordinate system is obtained as
\begin{eqnarray}
 \frac{d {\bm s}}{d t}
    \!\!&=&\!\!    {\bm s}
           \!\times\!
           \frac{e}{m}
           \Bigg(
                  (\frac{{\rm g_\mu}}{2}\!-\!1\!+\!\frac{1}{\tilde{\gamma}})
                  (1\!+\!3\epsilon^2\phi) {\bm B}
                  \nonumber\\
    & &        \hspace{12mm}
               -  (\frac{{\rm g_\mu}}{2}\!-\!\frac{\tilde{\gamma}}{\tilde{\gamma}+1})
                  (1\!-\!\epsilon^2\phi) {\bm \beta}\!\times\!{\bm E} 
                  \nonumber\\
    & &        \hspace{12mm}
               -  (\frac{{\rm g_\mu}}{2}\!-\!1)\frac{\tilde{\gamma}}{\tilde{\gamma}+1}
                  (1\!-\!\epsilon^2\phi) ({\bm \beta}\!\cdot\!{\bm B})\,{\bm \beta} 
           \Bigg)
           \nonumber\\
    &=&    {\bm s}
           \!\times\!
           \frac{e}{m}
           \Bigg(
                  \Big( \frac{{\rm g_\mu}}{2}\!-\!1\!+\!\frac{1}{\gamma}[1+2\epsilon^2\phi(\gamma^2\!-\!1)] \Big)
                  (1\!+\!3\epsilon^2\phi) {\bm B}
                  \nonumber\\
    & &        \hspace{12mm}
               -  \Big( \frac{{\rm g_\mu}}{2}\!-\!\frac{\gamma}{\gamma\!+\!1}[1\!-\!2\epsilon^2\phi(\gamma\!-\!1)] \Big)
                  (1\!-\!\epsilon^2\phi) {\bm \beta}\!\times\!{\bm E} 
                  \nonumber\\
    & &        \hspace{12mm}
               -  ( \frac{{\rm g_\mu}}{2}\!-\!1)\frac{\gamma}{\gamma\!+\!1}
                  [1\!-\!\epsilon^2\phi(2\gamma\!-\!1)] ({\bm \beta}\!\cdot\!{\bm B})\,{\bm \beta} 
           \Bigg)
           \nonumber\\
    &=&    {\bm \Omega}_{\rm s}^{\rm eff} \times {\bm s}
\label{eq: spin motion at inertial-frame in curved spacetime w/o LLI}
\end{eqnarray}
Putting $a_\mu\equiv {\rm g}_\mu/2\!-\!1$ and using
the effective value of the spin precession frequency in the curved spacetime (Eq.~(\ref{eq: spin motion at inertial-frame in curved spacetime w/o LLI}))
and
the effective value of the cyclotron frequency in the curved spacetime (Eq.~(\ref{eq: Omega_c by GR})),
the effective value of the anomalous spin precession frequency ${\bm \Omega}_{\rm a}^{\rm eff}$ is obtained as
\begin{eqnarray}
 {\bm \Omega}_{\rm a}^{\rm eff}
 &=&
  {\bm \Omega}_{\rm s}^{\rm eff}
 -{\bm \Omega}_{\rm c}^{\rm eff}
    \nonumber\\
 &=&
     \!- \frac{e}{m} 
       \Bigg(
            (1\!+\!3\epsilon^2\phi)\, a_\mu {\bf B}
           \nonumber\\
 & & \hspace{10mm}
        -  \Big[ 
                  a_\mu \!-\!  \frac{1}{\gamma^2\!-\!1}  
                  \!-\!  \epsilon^2\phi \Big( 
                         4\!+\!a_\mu
                   \!+\! \frac{3}{\gamma^2\!-\!1} 
                                    \Big)
           \Big] \,\,
           {\bm \beta}\!\times\!{\bf E} \,\,\,\,
           \nonumber\\
 & & \hspace{10mm}
        -  a_\mu\, \frac{\gamma}{\gamma\!+\!1} 
            \Big( 1\!-\!\epsilon^2\phi(2\gamma\!-\!1) \Big)  
           ({\bm \beta}\!\cdot\!{\bf B})\,{\bm \beta} 
      \Bigg)
.
\label{eq: Omega_a by GR}
\end{eqnarray}
When the motion is confined on the horizontal plane, magnetic field is applied vertically and the momentum is chosen to cancel the contribution of electric field, the effective value of the anomalous magnetic moment of muon is interpreted as
\begin{eqnarray}
a_{\mu ({\rm EXP})}^{\rm eff}
        &=&    \frac{ { \Omega}_a^{\rm eff}  }{ { \Omega}_L^{\rm eff} - { \Omega}_a^{\rm eff} } 
.
\label{eq: a_mu_exp^conv on curved spacetime}
\end{eqnarray}
The Larmor precession frequency in the curved spacetime can be interpreted as the special solution of the BMT equation with $u^\mu$=$(u^0,{\bm 0})$ and $s^\mu$=$(0, {\bm s})$, which corresponds to the null translational motion, as
\begin{eqnarray}
{\bm \Omega}_{\rm L}^{\rm eff} 
	= - (1+3\epsilon^2\phi)\,\, \frac{\rm g_\mu}{2} \frac{e}{m} {\bm B}
	=   (1+3\epsilon^2\phi)\,\,   {\bm \Omega}_{\rm L}
.
\end{eqnarray}
Therefore, we obtain
\begin{eqnarray}
 a_{\mu ({\rm EXP})}^{\rm eff}
   &=&        \frac{ { \Omega}_a^{\rm eff}  }{ { \Omega}_L^{\rm eff} - { \Omega}_a^{\rm eff} } \nonumber\\
   &=&        \frac{ (1+3\epsilon^2\phi)\,\,  {\Omega}_a  }
                   { (1+3\epsilon^2\phi)\,\,  {\Omega}_{\rm L} - (1+3\epsilon^2\phi)\,\,   {\Omega}_a } \nonumber\\
   &=&        \frac{  {\Omega}_a  }{  {\Omega}_{\rm L} -  {\Omega}_a } \nonumber\\ 
   &=&        a_{\mu ({\rm EXP})}
,   
 \label{eq: a_mu_exp^conv on curved spacetime-2}
\end{eqnarray}
which show that the effective value of the anomalous magnetic moment in the curved spacetime equals to that in the flat spacetime up to the post-Newtonian order $O(\epsilon^2)$ of the general relativity.
Thus the effective values of cyclotron frequency, spin precession frequency and Larmor precession frequency have been derived based on the primitive consideration based on kinematical framework. 
However those effective values in curved spacetime are respectively different from those of values in the flat spacetime,  
the gravitational contribution is canceled in the ratio Eq.~(\ref{eq: a_mu_exp^conv on curved spacetime-2}) and the anomalous magnetic moment in the curved spacetime coincides with the case in the flat spacetime.

\subsection{Gravitational Influence to Storage Ring Experiment to Measure Muon g-2}
We have found that the gravitational influence to the magnetic moment of muons in the storage ring experiment is canceled in the ratio based on the kinematical consideration up to the post-Newtonian order $O(\epsilon^2)$ using the Schwarzschild metric if the muon momentum is chosen to satisfy
\begin{eqnarray}
      a_\mu -  \frac{1}{\gamma^2\!-\!1} =0 ,
\end{eqnarray}
where $a_\mu$ is the experimental value of the anomalous magnetic moment.
Consequently the comparison between the experimental and theoretical values remains valid as before.
However, the Eq.~(\ref{eq: Omega_a by GR}) implies that the the general relativity modifies the contribution of the ${\bm \beta}\!\times\!{\bm E}$ term.
The perfect cancellation of the ${\bm \beta}\!\times\!{\bm E}$ term corresponds to
\begin{eqnarray}
	a_\mu -  \frac{1}{\gamma^2\!-\!1}  - \epsilon^2\phi \Big(  4\!+\!a_\mu \!+\! \frac{3}{\gamma^2\!-\!1} \Big)
	\,\,\,\,\sim\,\,\,\,
	(a_\mu +  2.8\times 10^{-9}) - \frac{1}{\gamma^2\!-\!1}
	= 0
	,
\end{eqnarray}
which introduces an offset to the experimental value of the anomalous magnetic moment as large as $2.8\times 10^{-9}$.
Therefore, if the ${\bm \beta}\!\times\!{\bm E}$ term is perfectly canceled in the experiment, the $a_\mu +  2.8\times 10^{-9}$ is measured instead of the $a_\mu$ and the offset may be misidentified as a discrepancy between the theory and the experiment.
The general relativistic offset may be clarified in future experiments with improved accuracy.

\section{Conclusion}
The magnetic moment of free fermions to be observed using instruments fixed on the Earth's surface has been examined by evaluating the effective magnetic moment including the influence of the gravitational field up to the post-Newtonian order $O(1/c^2)$ adopting the Schwarzschild metric as the background spacetime.
We found that the magnetic moment of fermions are influenced by the spacetime curvature as ${\bm \mu}_{\rm m}^{\rm eff}= (1+3\phi/c^2) \,\,{\bm \mu}_{\rm m} $ commonly to cases of the minimal coupling, non-minimal coupling and their mixture.
In the same manner, the effective values of gyromagnetic ratio and the anomalous magnetic moment depends on the Earth's gravitational field as
${\rm g}^{\rm eff}    \simeq (1 \!+\! 3\phi/c^2) \,\, {\rm g} $, \,\, 
${a}^{\rm eff} \simeq a + 3 (1\!+\!a) \,\,\phi/c^2\,\, $,
which seems inconsistent with the precise agreement between the experimental values and the theoretical values calculated in the flat spacetime.

However, the experimental values of the anomalous magnetic moment were obtained as the ratio of the spin precession frequency and the cyclotron frequency in the Penning trap and the storage ring experiments, and the gravitational influence is canceled to recover the consistency between the experiment and the theory.
Proper treatment on the experimental offset of the anomalous magnetic moment measured in the storage ring method would be necessary to compare theoretical and experimental values taking into consideration the change of the cancellation condition of the electric field contribution.

\section*{Acknowledgment}
This work is supported in part by a Grant-in-Aid for Science Research from JSPS (No.17K05453 to T.F).

%%%%%%%%%%%%%%%%%%%%%%%%%%%%%%%%%%%%%%%%%%%%%%%%%%%%%%%%%%%%%%%%%%%%%%%%%%
%       Appendix 
%%%%%%%%%%%%%%%%%%%%%%%%%%%%%%%%%%%%%%%%%%%%%%%%%%%%%%%%%%%%%%%%%%%%%%%%%%
\appendix
\section{\label{app:gamma}$\gamma$-matrix}
The $\gamma$-matrices in a curved spacetime satisfy
\begin{eqnarray}
   \{\gamma^\mu, \gamma^\nu\} &=& \gamma^\mu\gamma^\nu+\gamma^\nu\gamma^\mu=2 g^{\mu\nu} I_4
   ,
   \nonumber\\
   \{\gamma_\mu, \gamma_\nu\} &=& \gamma_\mu\gamma_\nu+\gamma_\nu\gamma_\mu=2 g_{\mu\nu} I_4
   ,
   \label{eq:gamma-matrix-in-GR}
\end{eqnarray}
where $I_4$ is the $4\times4$ unit matrix, $g_{\mu\nu}$ the metric tensor of the curved spacetime.
In the flat spacetime, the above relation leads to
\begin{eqnarray}
   \{\gamma^{(\alpha)}, \gamma^{(\beta)}\} &=& \gamma^{(\alpha)}\gamma^{(\beta)}+\gamma^{(\beta)}\gamma^{(\alpha)}=2 \eta^{\alpha\beta} I_4, \nonumber\\
   \{\gamma_{(\alpha)}, \gamma_{(\beta)}\} &=& \gamma_{(\alpha)}\gamma_{(\beta)}+\gamma_{(\beta)}\gamma_{(\alpha)}=2 \eta_{\alpha\beta} I_4, 
\end{eqnarray}
where $\eta_{\alpha\beta}$ is the metric tensor of the flat spacetime (Minkowski metric).
We define the tetrad $e^{\mu}_{(\alpha)}$ satisfying the relation
\begin{equation}
g_{\mu\nu} \,\,e^\mu_{(\alpha)} e^\nu_{(\beta)} = \eta_{\alpha\beta} =
\left(
 \begin{array}{cccc}
    \epsilon^{-2} &   0   &   0   &  0  \\
    0             &  -1   &   0   &  0  \\
    0             &   0   &  -1   &  0  \\
    0             &   0   &   0   & -1
 \end{array}
\right).\,\,\,\,\,\,
\end{equation}
The relation between the $\gamma$-matrices in the curved spacetime and the flat spacetime can be written as
\begin{equation}
   \gamma^{\mu}=\gamma^{(\alpha)} \,\,e^\mu_{(\alpha)} ,\,\,\,\,\,
   \gamma_{\mu}=\gamma_{(\alpha)} \,\,e_\mu^{(\alpha)} 
   \label{eq:gamma-matrix-in-SR}
\end{equation}
Using the tetrad $e^{\mu}_{(\alpha)}$ and $\gamma$-matrices, the spin connection $\Gamma_\mu$ can be written as
\begin{eqnarray}
 {\Gamma}_\mu &=&  \frac{1}{4}\, g^{\lambda\nu} \, e_\lambda^{(\alpha)} \, \nabla_\mu \, e_\nu^{(\beta)} 
                  \,\,\frac{1}{2}\, \Big[ \gamma_{(\alpha)}, \gamma_{(\beta)} \Big]
.
\label{eq:spin-connection}
\end{eqnarray}

Here we show the explicit representation of $\gamma$-matrices in the flat spacetime using the standard Dirac representation as
\begin{eqnarray}
&&
  \gamma^{(0)} = {\epsilon} \beta = 
\epsilon
\left(
 \begin{array}{cc}
    I &  0 \\
    0 & -I 
 \end{array}
\right),\,\,\,
 \gamma^{(i)} =
\left(
 \begin{array}{cc}
    0        &    \sigma_i \\
 - \sigma_i  &           0 
 \end{array}
\right), \nonumber\\
&&
 \gamma_{(0)} = \frac{1}{\epsilon}
\left(
 \begin{array}{cc}
    I &  0 \\
    0 & -I 
 \end{array}
\right),\,\,\,\,\,
 \gamma_{(i)} =
\left(
 \begin{array}{cc}
    0        &   -\sigma_i \\
   \sigma_i  &           0 
 \end{array}
\right)
,
\end{eqnarray}
where $\sigma_i$ is the Pauli matrix satisfying
\begin{equation}
 \sigma_i \, \sigma_j = \delta_{ij} + i\, \epsilon_{ijk} \,\sigma_k
\end{equation}
and their explicit representation is
\begin{equation}
  \sigma_1  = 
\left(
 \begin{array}{cc}
    0 &  1 \\
    1 &  0 
 \end{array}
\right),\,\,\,\,
 \sigma_2 =
\left(
 \begin{array}{cc}
    0      &  -i \\
    i      &  0 
 \end{array}
\right),\,\,\,\,
 \sigma_3 =
\left(
 \begin{array}{cc}
    1  &   0 \\
    0  &  -1 
 \end{array}
\right),\,\,\,\,\,\,\,
\end{equation}
The $4\times4$ $\rho_i$ matrices are Dirac matrices satisfying
\begin{equation}
 \rho_i \, \rho_j = \delta_{ij} + i\, \epsilon_{ijk} \,\rho_k
\end{equation}
and their explicit representation is
\begin{equation}
  \rho_1  = 
\left(
 \begin{array}{cc}
    0 &  I \\
    I &  0 
 \end{array}
\right),\,\,\,\,
 \rho_2 =
\left(
 \begin{array}{cc}
    0      &  -iI \\
    iI     &    0 
 \end{array}
\right),\,\,\,\,
 \rho_3 =
\left(
 \begin{array}{cc}
    I  &   0 \\
    0  &  -I 
 \end{array}
\right),\,\,\,\,\,\,\,
\end{equation}
which leads to
\begin{equation}
{\bf \alpha}_i = \rho_1 \sigma_i = 
\left(
 \begin{array}{cc}
    0        &  \sigma_i \\
    \sigma_i &  0 
 \end{array}
\right), \,\,\,\,\,\,\,\,
%\end{equation}
%\begin{equation}
{\bf \beta} = \rho_3  = 
\left(
 \begin{array}{cc}
    I   &   0 \\
    0   &  -I 
 \end{array}
\right).
\end{equation}

Here we define the 2nd-rank tensor $\sigma^{(\alpha)(\beta)}$ as
\begin{eqnarray}
%\Bigg( 
 \sigma^{(\alpha)(\beta)} &=& \frac{i}{2}\,\big[\gamma^{(\alpha)}, \gamma^{(\beta)}\big] ,\,\,\,\,\,\,\,\,
 \gamma^{(\alpha)} \gamma^{(\beta)} = \eta^{\alpha\beta}- i \,\sigma^{(\alpha)(\beta)} 
%\Bigg)
%\nonumber\\
\end{eqnarray}
\begin{eqnarray}
 \sigma^{(0)(k)} &=& \frac{i}{2}\,\big[\gamma^{(0)}, \gamma^{(k)}\big] = i\epsilon\rho_1\sigma_k = i\epsilon\alpha_k \nonumber\\
 \sigma^{(i)(j)} &=& \frac{i}{2}\,\big[\gamma^{(i)}, \gamma^{(j)}\big] =  \epsilon^{ijk} \sigma_k
\end{eqnarray}
\begin{eqnarray}
 \sigma_{(0)(k)} &=& \frac{i}{2}\,\big[\gamma_{(0)}, \gamma_{(k)}\big] = -i\epsilon^{-1}\rho_1\sigma_k = -i\epsilon^{-1}\alpha_k \nonumber\\
 \sigma_{(i)(j)} &=& \frac{i}{2}\,\big[\gamma_{(i)}, \gamma_{(j)}\big] =  \epsilon_{ijk} \sigma_k
.
\end{eqnarray}
Using the tetrad $e^{(\alpha)}_{\mu}$, 2nd-rank tensors $\sigma^{\mu\nu}$ and $F_{\mu\nu}$ can be written as
\begin{eqnarray}
  \sigma^{\mu\nu} &=& e^{\mu}_{(\alpha)} \, e^{\nu}_{(\beta)} \sigma^{(\alpha)(\beta)} \nonumber\\
  F_{\mu\nu}      &=& e_{\mu}^{(\alpha)} \, e_{\nu}^{(\beta)} F_{(\alpha)(\beta)}      %\nonumber\\
\end{eqnarray}
and
\begin{eqnarray}
    \sigma^{\mu\nu} F_{\mu\nu} &=& \sigma^{(\alpha)(\beta)}\,F_{(\alpha)(\beta)}
    .
\end{eqnarray}

\section{\label{app:Schwarzschild-geometry}Geometrical values in Schwarzschild metric}
The Schwarzschild metric can be written as
\begin{eqnarray}
  ds^2   &=&    (1\!-\!\epsilon^2\frac{2GM}{r}) c^2 dt^2 
              - \frac{dr^2 }{(1\!-\!\epsilon^2\frac{2GM}{r})} 
              - r^2(d\theta^2+\sin^2\!\theta\, d{\varphi}^2)
\end{eqnarray}
in the spherical coordinate.
This can be rewritten, using the isotropic coordinate and expanding up to the post-Newtonian order $O(\epsilon^2)$, as
\begin{eqnarray}
  ds^2 
       &=& \epsilon^{-2} (1\!+\!\epsilon^2 2\phi+\epsilon^4 2\phi^2) \, dt^2 
          - (1\!-\!\epsilon^2 2\phi) (dx^2\!+\!dy^2\!+\!dz^2)
           \nonumber\\
      & &  \hspace{7mm}  + O(\epsilon^4)
          ,
\label{app_eq: Schwarzschild-metric}
\end{eqnarray}
where $\phi\!=\!-GM/r$ is the Earth's gravitational potential.
When the metric tensor $g_{\mu\nu}$ is uniquely defined, the Christoffel symbol $\Gamma^\lambda_{\mu\nu}$ and the tetrad $e^{\mu}_{(\alpha)}$  are also uniquely determined in the following.

\subsection{\label{app:gmn}Metric tensor $g_{\mu\nu}$}
Using the definition and the Eq.~(\ref{app_eq: Schwarzschild-metric}), the explicit expression of each component of the four metric tensor $g_{\mu\nu}$ and $g^{\mu\nu}$ is given 
up to the post-Newtonian order $O(\epsilon^2)$, as
\begin{eqnarray}
 g_{00} &=&   \epsilon^{-2}\Bigl( 1 + \epsilon^2 2\phi + \epsilon^4 2\phi^2  \Bigr)  \nonumber\\
 g_{01} &=&    g_{02} =  g_{03} =   0  \nonumber\\
 g_{ij} &=&   -(1-2\epsilon^2\phi)\,\, \delta_{ij}  \nonumber \\
              \nonumber \\
 g^{00} &=&   \epsilon^{2}\Big( 1 - 2\epsilon^2\phi + \epsilon^4  2\phi^2  \Big) \nonumber\\
 g^{01} &=&   g^{02}  =  g^{03} =   0 \nonumber\\
 g^{ij} &=&   -(1+2\epsilon^2\phi)\,\, \delta^{ij}  
\label{app_eq: gmn}
\end{eqnarray}

\subsection{\label{app:Christoffel_symbol}Christoffel symbol $\Gamma^\lambda_{\mu\nu}$}
The Christoffel symbols are defined as
\begin{equation}
 \Gamma^\lambda_{\mu\nu}  =  \frac{1}{2}g^{\lambda\kappa}(g_{\kappa\mu;\nu}+g_{\kappa\nu;\mu}-g_{\mu\nu;\kappa}) 
.
\end{equation}
Substituting Eq.~(\ref{app_eq: gmn}), we obtain the explicit expressions of Christoffel symbols 
up to the post-Newtonian order $O(\epsilon^2)$, as
\begin{eqnarray}
&& \Gamma^0_{00}=0 \nonumber\\
&& \Gamma^1_{00}=\phi_1\hspace{1.3cm}\Gamma^2_{00}=\phi_2\hspace{1.3cm}\Gamma^3_{00}=\phi_3 \nonumber\\
&& \Gamma^0_{01}=\epsilon^2\phi_1\hspace{1.cm}\Gamma^0_{02}=\epsilon^2\phi_2\hspace{1.cm}\Gamma^0_{03}=\epsilon^2\phi_3 \nonumber\\
&& \Gamma^i_{0j}=0  \nonumber\\
&& \Gamma^0_{ij}=0 \nonumber\\
&& \Gamma^1_{11}=-\epsilon^2\phi_1\hspace{0.7cm}\Gamma^2_{11}=\epsilon^2\phi_2\hspace{1.0cm}\Gamma^3_{11}=\epsilon^2\phi_3\ \nonumber\\
&& \Gamma^1_{12}=-\epsilon^2\phi_2\hspace{0.7cm}\Gamma^2_{12}=-\epsilon^2\phi_1\hspace{0.7cm}\Gamma^3_{12}=0\nonumber\\
&& \Gamma^1_{13}=-\epsilon^2\phi_3\hspace{0.7cm}\Gamma^2_{13}=0\hspace{1.5cm}\Gamma^3_{13}=-\epsilon^2\phi_1\nonumber\\
&& \Gamma^1_{22}= \epsilon^2\phi_1\hspace{1.0cm}\Gamma^2_{22}=-\epsilon^2\phi_2\hspace{0.7cm}\Gamma^3_{22}=\epsilon^2\phi_3\nonumber\\
&& \Gamma^1_{23}=0\hspace{1.5cm}\Gamma^2_{23}=-\epsilon^2\phi_3\hspace{0.7cm}\Gamma^3_{23}=-\epsilon^2\phi_2\nonumber\\
&& \Gamma^1_{33}=\epsilon^2\phi_1\hspace{1.0cm}\Gamma^2_{33}=\epsilon^2\phi_2\hspace{1.0cm}\Gamma^3_{33}=-\epsilon^2\phi_3
,
\label{app_eq: Christoffel_symbol}
\end{eqnarray}
where $\phi_i=\partial_i\phi=-\frac{\phi}{r^2}x_i$.

\subsection{\label{app:tetrad} Tetrad $e^\mu_{(\alpha)}$}
Using Eq.~(\ref{app_eq: gmn}) and Eq.~(\ref{app_eq: Christoffel_symbol}), the tetrad $e^\mu_{(a)}$ satisfying
\begin{equation}
g_{\mu\nu} \,\,e^\mu_{(a)} e^\nu_{(b)} = \eta_{(a)(b)} =
\left(
 \begin{array}{cccc}
    \epsilon^{-2} &   0   &   0   &  0  \\
    0             &  -1   &   0   &  0  \\
    0             &   0   &  -1   &  0  \\
    0             &   0   &   0   & -1
 \end{array}
\right),\,\,\,\,\,\,
\label{app_eq: def_tetrad}
\end{equation}
can be obtained 
up to the post-Newtonian order $O(\epsilon^2)$, as
\begin{eqnarray}
 e^0_{(0)} &=&  1 - \epsilon^2 \phi+\epsilon^4\frac{1}{2}\phi^2 \nonumber\\
 e^i_{(0)} &=&  e^0_{(i)} =  0  \nonumber\\
 e^i_{(j)} &=&  (1 + \epsilon^2 \phi) \delta^i_j \nonumber
\end{eqnarray}
\begin{eqnarray}
 e^{(0)}_0 &=&    1 + \epsilon^2 \phi +\epsilon^4\frac{1}{2}\phi^2 \nonumber\\
 e^{(i)}_0 &=&    e^{(0)}_i =    0  \nonumber\\
 e^{(i)}_j &=&    (1 - \epsilon^2 \phi) \delta^i_j \nonumber%\\
\end{eqnarray}
\begin{eqnarray}
 e^{(0)0} &=&    \epsilon^2 \Big(1-\epsilon^2 \phi +\epsilon^4\frac{1}{2}\phi^2\Big) \nonumber\\
 e^{(0)i} &=&    e^{(i)0} =    0  \nonumber\\
 e^{(i)j} &=&    - (1 + \epsilon^2 \phi) \delta^i_j
.
\label{app_eq: tetrad}
\end{eqnarray}

\section{\label{app:LLI} Local Inertial Frame }
Tetrad enables to define the local inertial frame in which the local Lorentz invariance (LLI) is satisfied.
The four vectors in the local inertial frame $x^{(\mu)}$ and in the general coordinate system $x^{\mu}$ are related with each other using the tetrad as
\begin{eqnarray}
x^{(a)} &\equiv&   e^{(a)}_{\mu} x^{\mu} = (\tilde{x}^{0}, \tilde{{\bm x}})
.
\end{eqnarray}
Using the representation of the tetrad up to the post-Newton order $O(\epsilon^2)$ shown in Eq.~(\ref{app_eq: tetrad}),
the physical quantities relevant in this paper can be expressed as
\begin{eqnarray}
\tilde{S}^0       &=&  (1+\epsilon^2\phi) \, {S}^0     \nonumber\\
\tilde{\bm S}     &=&  (1-\epsilon^2\phi) \, {\bm S}   \nonumber\\
\tilde{\bm \beta} &=&  \frac{ \tilde{{\bm u}} }{ \tilde{u}^0  } 
                   =   (1-2\epsilon^2\phi) \,{{\bm \beta }}  \nonumber\\
\tilde{ t}        &=&  (1+\epsilon^2\phi) \,\,t    \nonumber\\
\tilde{\bm x}     &=&  (1-\epsilon^2\phi) \,\,{\bm x}    \nonumber\\
\tilde{ \gamma }  &=&  \Big( 1-2\epsilon^2\phi\, (\gamma^2-1) \Big) \,{{    \gamma}}   %\nonumber\\
\end{eqnarray}
The conversion of a tensor of rank 2 (electromagnetic tensor) is given as
\begin{eqnarray}
   F_{(a)(b)} &=& e_{(a)}^{\mu} e_{(a)}^{\nu}  F_{\mu\nu} %\nonumber\\ 
,
\end{eqnarray}
which results in
\begin{eqnarray}
 \tilde{\bm E}  &=&  F_{(0)(\alpha)}   
                 =   e_{(0)}^{0} e_{(\alpha)}^{\alpha}     F_{0\alpha} 
                 =   {\bm E}
                                      \nonumber\\ 
 \tilde{\bm B}  &=&  F_{(\alpha)(\beta)} 
                 =   e_{(\alpha)}^{\alpha} e_{(\beta)}^{\beta}  F_{\mu\nu} 
                 =   (1+2\epsilon^2\phi) {\bm B}
.
\end{eqnarray}

\section{\label{app:re-normalization} Normalization of Hamiltonian}
The scalar product of a spinor function $\Psi$ is defined as
\begin{equation}
\left\langle \Psi \middle\vert \Psi \right\rangle = \int \Psi^{\dagger} \Psi\, \sqrt{h}\, d^3x
,
\end{equation}
where $h_{ij}$ is the spatial components of the 4-dimensional tensor $g_{\mu\nu}$ : $h_{ij}=-g_{ij}$.
The expectation value of the hamiltonian is given as
\begin{equation}
\left\langle {\cal H} \right\rangle
=
\left\langle \Psi \middle\vert {\cal H} \middle\vert \Psi \right\rangle
=
\int \Psi^{\dagger}\, {\cal H} \,\Psi\, \sqrt{h}\, d^3x
.
\end{equation}
Here we put
\begin{eqnarray}
    \Psi' = h^{\frac{1}{4}}\, \Psi  
    ,\,\,\,\,\, \,\,\,\,\, 
    {\cal H}' = h^{\frac{1}{4}}\, {\cal H} \, h^{-\frac{1}{4}}\, 
\end{eqnarray}
to obtain
\begin{eqnarray}
\left\langle \Psi \middle\vert {\cal H} \middle\vert \Psi \right\rangle
&=& \int \Psi^{\dagger}\, {\cal H} \,\Psi\, \sqrt{h}\, d^3x \nonumber\\
&=& \int {\Psi'}^{\dagger}\, {\cal H}' \,{\Psi'}\, d^3x  \nonumber\\
&=& \left\langle \Psi' \middle\vert {\cal H}' \middle\vert \Psi' \right\rangle
.
\end{eqnarray}
This corresponds to the normalization of the expectation value of the hamiltonian into the expectation value in the local flat spacetime as ${\cal H}'$, which justifies that the $\epsilon$-dependent terms such as the post-Newtonian terms can be described as the perturbation to the terms in the flat spacetime
\cite{wajima-1997}.

\section{\label{app: typical scale of nabla-phi} Typical Scale of Gradient Terms of Gravitational Field }
In general the equation of translational motion of a charged particle in curved spacetime involving general relativistic effects can be written in Eq.~(\ref{eq: translational motion on curved space}). 
Using the Schwarzschild metric and considering up to the post-Newtonian order $O(\epsilon^2)$,   
the equation can be written in Eq.~(\ref{eq: translational motion on curved space-2}).
Here we estimate the magnitude of the contribution of the terms containing the gradient of the gravitational potential $\nabla\phi$ (last two terms in Eq.~(\ref{eq: translational motion on curved space-2})) in the precise measurement of the ${\rm g}_{\mu}\!-\!2$ in the storage ring apparatus.

The BNL E821 experiment \cite{muon-g-factor-BNL} has the apparatus parameters of
$\gamma$ = 29.3,
$B$ = 1.45 [T],
$r$ = 7112 [mm] and
$\Omega_{\rm c}  = 42  $ [MHz],
and the magnitudes of the electromagnetic interaction terms are estimated as
\begin{eqnarray}
    f_{\rm EM} 
      &=&  \Bigg|\, \Big( 1 \!+\! (2\gamma^2 \!+\! 1) \epsilon^2\phi \Big) \,\frac{e}{\gamma\,m} \,{\bm \beta}\!\times\!{\bm B} \,\Bigg| \nonumber\\
      &=&  \Big|\, \big( 1 \!+\! (2\gamma^2 \!+\! 1) \epsilon^2\phi   \big) \,\, \Omega_c \times \beta \,\Big| \nonumber\\
      &\simeq&  4.2 \times  10^{7} \,\,\,{\rm [s^{-1}] } 
.
\end{eqnarray}
On the other hands, the magnitude of terms containing the gradient of the gravitational potential is
\begin{eqnarray}
    f_{\nabla\phi} 
      &=&       \Big|\, \epsilon (1+\beta^2)\nabla\phi \Big|\, \nonumber\\
      &\simeq&  2 \times \frac{GM/R^2 }{c} \nonumber\\
      &\simeq&  6.5 \times  10^{-8} \,\,{\rm [s^{-1}] } %\nonumber\\
.
\end{eqnarray}
Therefore, the relative magnitude of the contributions of the gravity gradient and the electromagnetic interaction is
\begin{eqnarray}
  \frac{f_{\nabla\phi}}{f_{\rm EM}}  &\simeq& 1.5\times 10^{-15}
,
\end{eqnarray}
which shows the gravity gradient contribution is $10^{-15}$ times smaller than the electromagnetic interaction and is even $10^{-5}$ times smaller than the post-Newtonian effects.
Consequently, the contribution of the gravity gradient is negligible as long as the magnitude of the relevant contribution is $10^{-10}$ times of the main contribution.

%%%%%%%%%%%%%%%%%%%%%%%%%%%%%%%%%%%%%%%%%%%%%%%%%%%%%%%%%%%%%%%%%%%%%%%%%%%%%%%%%%%%%%%%%%%%%%%%%%%%%%%
%  References
%%%%%%%%%%%%%%%%%%%%%%%%%%%%%%%%%%%%%%%%%%%%%%%%%%%%%%%%%%%%%%%%%%%%%%%%%%%%%%%%%%%%%%%%%%%%%%%%%%%%%%%


\begin{thebibliography}{}

\bibitem[itzykson(1985)]{itzykson-1985}
	             C. Itzykson and J. B. Zuber, 
                     {\it Quantum Field Theory}, McGraw-Hill, New York (1985)

\bibitem[peskin(1997)]{peskin-1997}
	             M. E. Peskin and D. V. Schroeder, 
                     {\it An Introduction to Quantum Field Theory}, Addison-Wesley (1997)

\bibitem{anandan-1984}
	             J. Anandan, Phys.\ Rev.\ D {\bf 30}, 1615 (1984).
\bibitem{parker-1980}
	             L. Parker, Phys.\ Rev.\ D, {\bf 22}, 1922 (1980)
\bibitem{wajima-1997}
           	     S. Wajima, M. Kasai and T. Futamase, Phys.\ Rev.\ D, {\bf 55}, 1964 (1997) 

\bibitem[tani(1951)]{FWT}
                     L. L. Foldy and S. A. Wouthuysen, Phys.\ Rev., {\bf 78}, 29 (1950);
                     S. Tani, Prog.\ Theor.\ Phys., {\bf 6}, 267 (1951);
	             K. M. Case, Phys.\ Rev., {\bf 95}, 1323 (1954)

\bibitem[lightman(1973)]{lightman-1973}
                     %Restricted proof that the weak equivalence principle implies the Einstein equivalence principle, 
                     A. P. Lightman  and D. L. Lee,  Phys.\ Rev.\ D, {\bf 8}, 364, (1973)

\bibitem[will(1974)]{will-1974}
                     % gravitationally modified magnetic moment
		     C. M. Will, Phys.\ Rev.\ D, {\bf 10}, 2330 (1974)


\bibitem[will(1993)]{PPN}
		     C. M. Will, {\it Theory and experiment in gravitational physics}, Cambridge (1993) 

\bibitem{Aoyama-2015}
	              % electron g-2 with 10-th order correction
	              T. Aoyama, M. Hayakawa, T. Kinoshita and M. Nio, Phys.\ Rev.\ D {\bf 91}, 033006 (2015)


\bibitem{electron-g-2-experiment} 
	              D. Hanneke, S. Fogwell, and G. Gabrielse, Phys.\ Rev.\ Lett., {\bf 100}, 120801 (2008);
	              D. Hanneke, S. Fogwell Hoogerheide, and G. Gabrielse, Phys.\ Rev.\ A {\bf 83}, 052122 (2011).

\bibitem{CODATA-2014}
	              P. J. Mohr, D. B. Newell, and B. N. Taylor, Rev.\ Mod.\ Phys. {\bf 88}, 035009, (2016)

\bibitem{PDG-2016}
	              C. Patrignani et al. (Particle Data Group), Chin.\ Phys.\ C, {\bf 40}, 100001 (2016)

\bibitem{muon-g-factor-CERN} 
		      %review of muon g-2 values experiment in CERN
		      J. Bailey et al., Nucl.\ Phys.\ B, {\bf 150}, 1 (1979); 
		      J. Bailey et al., Phys.\ Lett. B, {\bf 68}, 191 (1977)

\bibitem{muon-g-factor-BNL} 
		      %muon g-2 values experiment in E821 BNL  
		      G. W. Bennett {et al.}, Phys.\ Rev.\ D, {\bf 73}, 072003 (2006);
	              J. P. Miller {et al.}, Rep. Prog. Phys. 70, 795 (2007);
	              J. M. Paley, ``Measurement of the Anomalous Magnetic Moment of the Negative Muon to 0.7 Parts Per Million'', Ph.D. Theses, Boston University (2004)

\bibitem{BMT-equation} 
		      %relativistic estimation of muon g-2 values 
		      V. Bargmann, L. Michel and  V. L. Telegdi, Phys.\	Rev.\ Lett., {\bf 2}, 435 (1959)


\bibitem[jackson(1998)]{jackson-1998}
		      J. D. Jackson, {\it Classical Electrodynamics (3rd edition)}, Wiley (1998)


\end{thebibliography}
\end{document}